\documentclass[a4paper,11pt,fleqn]{article}
\pdfoutput=1 
\usepackage[utf8]{inputenc}
\usepackage{jheppub} 

\usepackage{amssymb,amsmath,amsfonts}
\usepackage{mathtools}
\usepackage{graphicx}
\usepackage{multirow}
\usepackage{verbatim}
\usepackage{appendix}
\usepackage{fancybox}
\usepackage{array}
\usepackage{url}
\usepackage{float}
\usepackage[mathscr]{euscript}
\usepackage{bbm} 
\usepackage{slashed}
\usepackage{tikz}                     
\usepackage{tikz-cd}

\newcommand{\Z}{{\mathbb Z}}
\newcommand{\R}{{\mathbb R}}
\newcommand{\C}{{\mathbb C}}
\newcommand{\Q}{{\mathbb Q}}
\newcommand{\RP}{{\mathbb{RP}}}

\def\CH{{\mathcal H}}

\def\CN{{\mathcal N}}

\def\Tr{{\mathrm{Tr}\,}}

\def\Ker{{\mathrm{Ker}\,}}
\def\Hom{{\mathrm{Hom}\,}}
\def\rank{{\mathrm{rank}\,}}
\def\pt{{\mathrm{pt}}}

\def\spinor{{\mathscr{S}}}
\def\SUB{{\mathfrak{S}}}
\def\tilde{\widetilde}
\def\hat{\widehat}
\def\bar{\overline}

\title{Trialities of minimally supersymmetric 2d gauge theories}

\author[1,2]{Sergei Gukov}
\author[1,3]{Du Pei}
\author[4]{Pavel Putrov}

\affiliation[1]{Walter Burke Institute for Theoretical Physics, California Institute of Technology, Pasadena, CA 91125, USA}
\affiliation[2]{Max-Planck-Institut f\"{u}r Mathematik, Vivatsgasse 7, D-53111 Bonn, Germany}
\affiliation[3]{Mathematical Sciences Research Institute, 17 Gauss Way, Berkeley 94720, USA}
\affiliation[4]{ICTP, Strada Costiera 11, Trieste 34151, Italy}

\preprint{CALT-TH-2019-047}

\abstract{We study dynamics of two-dimensional $\mathcal{N}=(0,1)$ supersymmetric gauge theories. In particular, we propose that there is an infrared \textit{triality} between certain triples of theories with orthogonal and symplectic gauge groups. The proposal is supported by matching of anomalies and elliptic genera. This triality can be viewed as a $(0,1)$ counterpart of the $(0,2)$ triality proposed earlier by two of the authors and A.~Gadde. We also describe the relation between global anomalies in gauge theoretic and sigma-model descriptions, filling in a gap in the present literature.}

\begin{document}
\maketitle

\section{Introduction}
A direct analysis of interacting quantum field theories in the strongly coupled quantum regime is often quite difficult due to absence of a good universal approach to calculating the path integral beyond its perturbative expansion. Alternatively, one can construct quantities protected under continuous deformations, including renormalization group flow, and use them to constrain the dynamics of the theory in the infrared. An important class of such protected quantities is given by 't Hooft anomalies of global symmetries. Supersymmetry allows construction of additional protected observables that can provide a refined information about the infrared dynamics. This approach has proved to be very fruitful in different dimensions and for various amount of supersymmetry. In two dimensions, most of the known results so far rely on the existence of at least $\CN=(0,2)$ supersymmetry (one of the seminal papers is \cite{Witten:1993yc}). Interesting results on $\CN=(0,1)$ non-linear sigma models (NLSMs) were obtained recently in \cite{Gaiotto:2019asa,gaiotto2019mock}. 

In this paper we provide exact results on $\CN=(0,1)$ (non-abealian) gauged linear sigma models (GLSMs). In particular we discrover \textit{trialities} between certain $\CN=(0,1)$ GLSMs which are very similar in nature to $\CN=(0,2)$ trialities studied in \cite{Gadde:2013lxa,Gadde:2014ppa}. In a certain sense the $\CN=(0,1)$ theories participating in the trialities can be understood as ``real slices'' of their ``complex'' $\CN=(0,2)$ counterparts, both on the level of gauge theory description and effective non-linear sigma model. In particular, the target spaces of NLSMs are real and complex Grassmannians respectively. It would be interesting to formulate such a $(0,2)$/$(0,1)$ correspondence more generally.  

It is worth noting that many methods that have been successfully used to analyze $\CN=(0,2)$ GLSMs, such as localization \cite{Gadde:2013ftv,Benini:2013xpa} and c-extremization \cite{Benini:2012cz}, are not available in the $\CN=(0,1)$ setting. Nevertheless, $\CN=(0,1)$ theories have well-defined elliptic genus \cite{Witten:1986bf} that, as we demonstrate in this paper, can still be computed for at least certain GLSMs using their effective NLSM description and the Atiyah--Bott localization formula.

2d $(0,1)$ theories have a surprising connection with stable homotopy theory, and are believed to represent cocycles in a generalized cohomology theory known as TMF (see \cite{douglas2014topological} for a comprehensive review of this subject). This relation was first proposed by Stolz and Teichner \cite{stolz2004elliptic,stolz2011supersymmetric}, based on earlier work of Segal \cite{segal2007elliptic}, and implies that there is a completely new set of invariants of 2d $(0,1)$ theories that take values in the ring of ``Topological Modular Forms.'' This new set of invariants refine the elliptic genus, and are expected to be \emph{complete}. In other words, they uniquely characterize deformation classes of 2d $(0,1)$ theories. These new invariants was recently studied from the physics point of view in \cite{gaiotto1811holomorphic, Gaiotto:2019asa,gaiotto2019mock}, and in \cite{gukov20184} for their connection with the topology of 4-manifolds. 

It is expected from physics that, once we consider 2d $(0,1)$ theories with a flavor symmetry $G$, there is a even finer set of invariants --- ``$G$-equivariant Topological Modular Forms'' --- that refines the flavored elliptic genus and uniquely characterize classes of theories under deformations preserving both supersymmetry and the flavor symmetry $G$. Some preliminary discussions about their properties can be found in \cite[Sec.~3.2]{gukov20184}, but at this stage, a mathematical theory for them is still lacking. The present work provides one more motivation for developing the equivariant theory of TMF. Indeed, the unflavored elliptic genera of almost all theories considered in Section~\ref{sec:SO-GLSMs} of this paper are identically zero, while the flavored versions contain rich information about the theories. One also expects similar phenomenon to happen for their refined counterparts --- the $G$-equivariant Topological Modular Forms would become trivial in the ``unflavored limit.''

Another motivation to compare dualities with different amounts of supersymmetry is that, sometimes, such relations can be very illuminating, or even lead to new dualities. For example, recently, this approach was successfully used in three-dimensions \cite{Kachru:2016rui,Kachru:2016aon} where, starting with a well established mirror symmetry of 3d $\CN=4$ gauge theories \cite{Intriligator:1996ex}, one can consider a gradual cascade of soft supersymmetry breaking to derive $\CN=2$ dualities \cite{Aharony:1997bx}, or even $\CN=0$, non-supersymmetric particle-vortex dualities~\cite{Karch:2016sxi,Seiberg:2016gmd,Murugan:2016zal}. It would be interesting to explore whether the first examples of dualities in non-abelian 2d $(0,1)$ gauge theories proposed here, either in compact or in non-compact models, can be related in a similar way to 2d $(0,2)$ dualities or, possibly, $(0,4)$ dualities.

The paper is organized as follows. In Section \ref{sec:01-SUSY}, we provide a brief overview of $\CN=(0,1)$ supersymmetry in 2d and give general comments on the correspondence between (both global and perturbative) anomalies and topological terms in GLSM and NLSM descriptions. In Section \ref{sec:SO-GLSMs}, we consider in detail a family of $\CN=(0,1)$ GLSMs with $SO(n)$ gauge groups and provide evidence for trialities between theories from this family. In Section \ref{sec:generalizations}, we consider various generalizations. In Appendix \ref{app:coh-grassmannian}, we provide useful facts and formulas on cohomology of real Grassmannians.

\section{$\CN = (0,1)$ supersymmetry in 2d}
\label{sec:01-SUSY}

\subsection{$\CN = (0,1)$ superspace and supermultiplets}

The $\CN=(0,1)$ super-Poincar\'e symmetry contains a single \textit{real} supercharge $Q_+$ of positive chirality which squares to the difference between the Hamiltonian $H$ and the momentum $P$ (i.e.~time and spatial translations):
\begin{equation}
    Q^2_+=H-P
\end{equation}
Unlike the better studied case of $(0,2)$ or $(2,2)$ supersymmetry, there is no (continuous) $R$-symmetry. In a sense, there is only $\Z_2$ R-symmetry which can be identified with fermionic parity acting on the right-moving sector. 

The $(0,1)$ superspace has local coordinates $(x^+,x^-,\theta^+)$, where $x^{\pm}$ are the standard light-cone coordinates and $\theta^+$ is a single (self-conjugate) Grassmann coordinate. As usual, it is customary to introduce the corresponding derivatives $\partial_\pm:=\partial/\partial x^\pm$, $D_+:=\partial/\partial \theta^++i\theta^+\partial_+$.

For the theories without gravity, there are three basic supermultiplet of $(0,1)$ super-Poincar\'e symmetry: scalar, Fermi and vector \cite{Sakamoto:1984zk,Hull:1985jv,Brooks:1986uh}.

\begin{itemize}
    \item \textit{Fermi multiplet}
    \begin{equation}
    \Gamma(x,\theta)=\psi_-(x)+\theta^+F(x) 
\end{equation}
where $\psi_-$ is a left-moving Majorana--Weyl spinor field and $F$ is an auxiallary scalar field.

    \item \textit{Scalar multiplet}
    \begin{equation}
    \Phi(x,\theta)=\phi(x)+\theta^+ \psi_+(x) 
\end{equation}
where $\psi_+$ is a right-moving Majorana--Weyl spinor field and $\phi$ is a scalar field.

    \item \textit{Vector multiplet} (after gauge fixing fermionic components)
\begin{equation}
\begin{array}{rl}
     \Lambda_+(x,\theta) & = \theta^+A_+(x)  \\
     \Lambda_-(x,\theta) &  = A_-(x)+\theta^+\lambda_-(x)
\end{array}    
\end{equation}
where $\lambda_-$ is a left-moving Majorana--Weyl spinor field in the adjoint representation of the gauge group and $A_\pm$ are the light-cone coordinates of a gauge field, which is locally an adjoint valued one-form. It is also useful to introduce the corresponding covariant derivatives
\begin{equation}
\begin{array}{rl}
     \nabla_+ & = D_++i\Lambda_+  \\
     \nabla_- &  = \partial_-+i\Lambda_-
\end{array}    
\end{equation}
and the super field strength
\begin{equation}
    \Sigma:=i[\nabla_+,\nabla_-]=\lambda_-+\theta^+ F_{+-}
\end{equation}
which has the structure of a Fermi multiplet in the adjoint representation of the gauge group, with the ordinary field strength $F_{+-}$ playing the role of the auxiliary scalar field. 

\end{itemize}

\subsection{Two-dimensional $\CN = (0,1)$ SQCD}
\label{sec:GLSM-review}

In this paper we are primarily interested in dynamics and dualities of 2d gauge theories with non-abelian gauge group and minimal supersymmetry. Naturally, a theory of this type could be called either a supersymmetric QCD (SQCD) in two dimensions or a gauged linear sigma-models (GLSM).\footnote{For balance, we will use both names interchangeably.} Specifically, our 2d $(0,1)$ SQCD or GLSM models are labeled by the following data: 
\begin{itemize}
    \item Number of scalar and Fermi multiplets, $n_b\in \Z_{\ge 0}$ and $n_f\in \Z_{\ge 0}$, respectively. Equivalently, the scalar and Fermi multiplets are chosen to be valued in real vector spaces $\R^{n_b}$ and $\R^{n_f}$ equipped with standard bilinear pairing.\footnote{General vector spaces with a generic non-degenerate pairing can be reduced to this case by field redefinition.}
    \item A compact (not necessarily connected) gauge Lie group $G$ together with homomorphisms $\rho_{b}:G\rightarrow O(n_{b})$ and $\rho_{f}:G\rightarrow O(n_{f})$. Equivalently, there is a choice of representations of $G$ preserving the bilinear pairings on the vector spaces where scalar and Fermi multiplets take their values.
    \item A $G$-equivariant map $J:\R^{n_b}\rightarrow \R^{n_f}$, known as the $(0,1)$ superpotential.
\end{itemize}
In terms of this data, the action then reads
\begin{multline}
    S_\text{SQCD}=
    \int d^2xd\theta^+\left(
    \frac{i}{2}\sum_{i=1}^{n_b}\nabla_+\Phi^i \nabla_- \Phi^i
    -\frac{1}{2}\sum_{a=1}^{n_f}\Gamma^a \nabla_+ \Gamma^a+\right.\\
    \left.
    +\frac{1}{2g^2}\langle \Sigma,\nabla_+ \Sigma \rangle
    +m\sum_{a=1}^{n_f}\Gamma^a J^a(\Phi)
    \right)+S_\text{top}
\end{multline}
where the action of the covariant (super) derivatives $\nabla_\pm$ on the scalar and Fermi multiplets is determined by the homomorphisms $\rho_b$ and $\rho_f$ respectively. The bracket $\langle \cdot , \cdot \rangle$ denotes the Killing form on the Lie algebra $\frak{g}:=\mathrm{Lie}(G)$. The parameters $g$ and $m$ have the dimension of mass. With this convention all the coefficients in the parameters in the functions $J^a(\Phi)$ are dimensionless, since $\Phi^i$ are dimensionless.

Integrating the first three terms over $\theta^+$ produces the standard kinetic terms for the component fields as well as their coupling to the gauge field $A_\pm$. The last term, after integrating out the auxiliary fields $F^a$ in the path integral produces the supersymmetric combination of Yukawa couplings and a scalar potential:
\begin{equation}
    \int d^2xd\theta^+\,m\sum_{a=1}^{n_f}\Gamma^a J^a(\Phi)
   \quad \overset{\int DF}{\rightsquigarrow} \quad
        \int d^2x\left(
    m\sum_{i,a}\psi_+^i\psi_-^a\frac{\partial J^a}{\partial \Phi^i}(\phi)
    +\frac{m^2}{2}\sum_{a}(J^a(\phi))^2
    \right).
\end{equation}

The choice of the topological term $S_\text{top}$ is determined by the generalized theta-angle \cite{Kapustin:2014dxa} (alternatively it can be described by the so-called supercohomology \cite{Wang:2017moj,Wang:2018pdc}, see also \cite{Gaiotto:2015zta,brumfiel2016pontrjagin}),
\begin{equation}
    \alpha\in \Hom\left(\Omega_2^\text{Spin}(BG),2\pi \R/\Z\right),
\end{equation}
and can be written as 
\begin{equation}
    S_\text{top}=\alpha([(\Sigma,f_G)])
    \label{GLSM-top-term}
\end{equation}
where $\Sigma$ is the world-sheet, understood as a spin 2-manifold, and $f_G:\Sigma\rightarrow BG$ is the map (considered up to homotopy) which determines the isomorphism class of the gauge bundle. Together, the pair $(\Sigma,f_G)$, by definition represents an element of the spin bordism group $\Omega_2^\text{Spin}(BG)$. The addition of the topological term can be understood as a coupling of the gauge theory to a fermionic SPT with $G$ symmetry.

At the quantum level, due to the presence of chiral fermions, the theory in general can suffer from gauge anomalies. This gives a restriction on the allowed representations $\rho_b$ and $\rho_f$. When the Lie algebra $\frak{g}$ is simple, the cancellation condition on the \textit{perturbative} anomalies, as usual, can be given in terms of the Dynkin indices of the representations $\rho_{b}$ and $\rho_{f}$, understood as the representations of $\frak{g}$:
\begin{equation}
T(\rho_b)-T(\rho_f)-T(\frak{g}) \;= \; 0 \,.
\label{TTTanom}
\end{equation}
As a reminder, the Dynkin index $T(\rho)$ of an \textit{irreducible} representation $\rho$ is related to its quadratic Casimir $C_2(\rho)$ as
\begin{equation}
    T(\rho)=C_2(\rho)\,\frac{\dim(\rho)}{\dim(\frak{g})}.
\end{equation}

The cancellation condition of the global anomalies (for which the global structure of $G$ becomes relevant) is more subtle and will not be addressed here in full generality. Briefly, it is required that a certain element (determined again by the representations $\rho_{b,f}$) in the finite abelian group $\Hom\big(\text{Tor}\,\Omega_3^\text{Spin}(BG),U(1)\big)$ vanishes. Instead, we will check that this condition holds for the particular theories that will be considered in this paper.

The theory also has the gravitational anomaly 
\begin{equation}
    c_R-c_L=\frac{1}{2}(n_b-n_f-\dim G).
    \label{gravanom}
\end{equation}

\subsection{Sigma-models}
\label{sec:sigma-models}

At the \textit{classical} level, the $(0,1)$ non-linear sigma-model (NLSM) is specified by the following data:

\begin{itemize}
    \item A Riemannian manifold $(X,g)$ called the \textit{target space}.
    \item A smooth vector bundle $\pi:E\rightarrow X$ equipped with an inner product $h(p):E_p\otimes E_p \rightarrow \R$ on the fibers $E_p=\pi^{-1}(p),\;p\in X$, and a connection, locally specified by the connection 1-form $A$ on $X$. Informally, $E$ can be called \textit{a bundle of left-moving fermions}. 
    \item A closed 3-form $H\in \Omega^3(X),\; dH=0$, such that $\int_{\Xi} H\in \Z$ for any closed 3-cycle $\Xi$ in $X$.  The 3-form $H$ is also known as \textit{NS-NS flux} and locally is given by $H=dB$ where $B$ is the Kalb--Ramond 2-form field.\footnote{To be precise, $B$ is a connection on a gerbe on $X$ and $H$ should be considered as an element of differential cohomology of $X$ with integral coefficients in degree 3, the natural home for the curvature of a connection on a gerbe on $X$.}
\end{itemize}

In terms of this data, the action of the sigma-model with a world-sheet $\Sigma$ takes the following form \cite{Hull:1985jv} (in order to make the expression lighter, we assume in the formula below that $\Sigma$ itself is flat; if needed, the coupling to the metric on $\Sigma$ can be easily restored)
\begin{multline}
    S_\sigma= \\ 
    \int_\Sigma d^2x d\theta^+\left(\sum_{i,j}g_{ij}(\Phi)D_+\Phi^i\partial_-\Phi^j
    +\sum_{a,b}h_{ab}(\Phi)\Gamma^a \left(D_++\sum_iA^a_{b,i}D_+\Phi^i\right) \Gamma^b
    \right)\\
   +2\pi\int_{\Xi} \phi^*(H)-2\pi i\int_{\Sigma} d^2x\, H_{ijk}\psi_+^i\psi_+^j\partial_-\phi^k
\end{multline}
where $\Phi^i$ are components of the map $\phi:\Sigma\rightarrow X$ promoted to scalar superfields and $\Gamma^a$ are components of a Fermi superfield valued in the fiber $E_{\phi(x)}$. The last two terms are the supersymmetrization of the Wess--Zumino--Witten term. The bosonic part is written in terms of a 3-manifold $\Xi$ such that $\partial \Xi=\Sigma$ and the map $\phi$ is extended to a map $\Xi \rightarrow X$.\footnote{In general such extension can be obstructed. The proper way to define the WZW term in general is as the holonomy of the gerbe connection (locally given by the Kalb-Ramond 2-form field $B$) along $\Sigma$.}

As in the GLSM, at the quantum level the theory generically suffers from the anomalies. In particular, integrating out chiral fermions in general does not produce a well-defined function on the space of maps $\Sigma\rightarrow X$. The first condition for the anomaly cancellation is that $w_1(TX)=w_1(E)$, i.e.~that $TX\oplus E$ is orientable. This condition can be already seen at the level of reduction to a supersymmetric matrix model, where the path integral becomes just the ordinary integral over $X$ of the Grassmann integral over the fiber of $E$. This will have a globally well-defined sign if $w_1(TX)=w_1(E)$. The second condition is the requirement that there is a globally defined bundle $\spinor TX\otimes \spinor E$, where $\spinor V$ denotes a spinor bundle of the real vector bundle $V$ equipped with an inner product. The obstruction to the existence of such bundle globally is given by $w_2(TX)=w_2(E)\in H^2(X,\Z_2)$. Note that the individual spinor bundles $\spinor TX$ and $\spinor E$ may not exist. The global choice of $\spinor TX\otimes \spinor E$ is in general not unique (when $E$ is trivial this choice is equivalent to the choice of spin structure on $X$) and the quantum theory depends on it.  This anomaly can be already seen at the level of $\CN=1$ supersymmetric quantum mechanics obtained by dimensional reduction from 2d to 1d. In particular, the Hilbert space of such quantum mechanics is the space of (square-integrable) sections of $\spinor TX\otimes \spinor E$ and the supercharge is the Dirac operator, which can be well defined only when $w_2(TX)=w_2(E)$. The third condition on the topology of $X$ and $E$ is
\begin{equation}
    \frac{1}{2}(p_1(X)-p_1(E))=0\in H^4(X,\Z)
    \label{string-condition}
\end{equation}
and cannot be seen just at the level of quantum mechanics. Roughly, it comes from the condition that there is a well-defined Dirac operator on the \textit{loop space} of $X$.  Note that the $1/2$ operation on the left-hand side in the formula above is well defined due to the first condition $w_2(TX)-w_2(E)=0$ and the fact that $p_1=w_2^2\bmod 2$. The corresponding anomaly can be also understood as the anomaly with respect to diffeomorphisms of the target space $X$ and gauge transformations of the connections on the bundle $E$. As usual, this anomaly can be seen at one loop. The invariance under diffeomorphisms and gauge transformations can be cured by modifying the gauge transformation property of the Kalb--Ramond field $B$ \cite{Hull:1985jv,Hull:1986xn,Howe:1987nw}. This, however, modifies the usual Bianchi identity ($dH=0$) for the corresponding field strength
\begin{equation}
    {dH}=-\frac{1}{16\pi^2}\Tr F_{TX}\wedge F_{TX} +\frac{1}{16\pi^2}\Tr F_E\wedge F_E 
    \label{bianchi-modified}
\end{equation}
where $F_{TX}$ is the curvature of a certain connection on $TX$ (obtained by shifting the Levi-Civita connection by $H$) and $F_E$ is the curvature of the 1-form connection $A$ on $E$. By passing to the de Rham cohomology, one gets (\ref{string-condition}) but in $H^4(X,\C)$. When $E$ is trivial, the choice of the 3-form $H$ trivializing the difference of the Chern--Weil realizations of the Pontryagin classes in (\ref{bianchi-modified}) is called a \textit{geometric string structure} of $X$ (the full data of the geometric string structure is more involved, see e.g.~\cite{2009arXiv0906.0117W} for details). The geometric string structure is a refinement of a \textit{topological string structure}, that is a choice of a homotopy class of a lift 
\begin{equation}
\begin{tikzcd}
 & BString \ar[d] &  
\\
 X \ar[ur,dashed] \ar[r] & BSpin \ar[r,"p_1/2"] & K(\Z,4)
\end{tikzcd}
\end{equation} where $BSpin$ is the classifying space of (stable) Spin bundles. 
The space of topological string structures is non-canocially isomorphic to $H^3(X,\Z)$; the ambiguty can be understood as a shift of $H$ by a curvature of gerbe, modulo an exact form.

\subsection{Flowing from gauge theory to sigma-model}
\label{sec:RGflow}

Let us assume that the system of equations $J^a(\phi)=0$ given by the superpotential of a GLSM is non-degenerate and the action of the gauge group on the corresponding space of solutions $Y:=\{J^a(\phi)=0\}$ is free. On the classical level, the parameters in the functions $J^a(\phi)$ are marginal while $g,m\rightarrow \infty$ under RG flow. In the limit $m\rightarrow \infty$ the values of the scalar fields $\phi^i$ are restricted to $Y:=\{J^a(\phi)=0\}$. Moreover, in the limit $g\rightarrow \infty$ only gauge invariant combinations of the fields survive. Therefore, under some approximation\footnote{Meaning the UV scale $\Lambda\ll m,g$. But what is important, is that independently of this condition, the GLSM and NLSM can be continuously connected in the space of $\CN=(0,1)$ quantum field theories and therefore the quantities protected under continuous deformations remain the same.}, before reaching the IR fixed point the GLSM flows to the sigma-model with target space
\begin{equation}
    X=Y/G
    \label{XYGtarget}
\end{equation}
and the bundle of left-moving fermions
\begin{equation}
 E=\Ker \frac{\partial J}{\partial \phi}\;   
 \label{bundle-from-GLSM}
\end{equation}
where $\frac{\partial J}{\partial \phi}$ is understood as a map $Y\times \R^{n_f}\rightarrow \R^{n_b}$. As was mentioned in Section \ref{sec:sigma-models}, to define the theory at the quantum level it is also necessary to specify the choice of trivialization of $w_2(TM)-w_2(E)$ and $(p_1(E)-p_1(TX))/2$. Such trivializations are fixed by the way the manifold $X$ and the bundle $E$ are realized via the equations $J_a=0$ and also a choice of the topological term (\ref{GLSM-top-term}). Let us briefly mention the general situation and later give a more concrete correspondence for the family of gauge theories of interest. 

For the sake of technical simplicity of the argument, assume that $E$ itself is trivial. Then, such trivializations are the spin and string structure on $X$. The case of the non-trivial bundle is similar. The spin and string structure on $Y=\{J^a(\phi)=0\}$ are induced from the standard spin and string structure on $\R^{n_b}$ using the trivialization of the normal bundle given by the vectors $\partial{J^a}/\partial{\phi^i}$. Given that the $G$-action is non-anomalous, the spin and string structures on $Y$ induce those on $X=Y/G$. The sets of spin and string structures on $X=Y/G$ are torsors over $H^1(X,\Z_2)$ and $H^3(X,\Z)$ respectively. (The same is true for the trivializations in the case of non-trivial bundle $E$.) The cohomology of the quotient with coefficients in $R$ can be calculated using Cartan--Leray spectral sequence
\begin{equation}
 H^p(BG,H^q(Y,R))\Rightarrow H^{p+q}(Y/G,R).   
\end{equation}
In particular, there are universal (i.e.~independent of $Y$ itself) contributions $H^1(BG,\Z_2)$ and $H^3(BG,\Z)$ to the cohomologies classifying, respectively, spin and string structures on $X=Y/G$. On the other hand, there 
is an Atiyah--Hirzerbruch spectral sequence
\begin{equation}
    E_2^{p,q}=H^p\left(BG,\Hom\left(\Omega_q^\text{Spin}(\pt),2\pi \R/\Z\right)\right)\Rightarrow \Hom\left(\Omega_{p+q}^\text{Spin}(BG),2\pi \R/\Z\right).
\end{equation}
The terms with $p+q=2$ are known to be stabilized already on the second page. Namely, there is a filtration
\begin{equation}
 \Hom\left(\Omega_{2}^\text{Spin}(BG),2\pi \R/\Z\right)  =F^3\supset F^2\supset F^1\supset F^0=0 
 \label{AH-filtration}
\end{equation}
with $F^{p+1}/F^{p}=E_2^{p,2-p}$, where
\begin{equation}
    \begin{array}{rl}
    E_2^{2,0}=& H^2(BG,2\pi \R/\Z), \\
    E_2^{1,1}=& H^1(BG,\Z_2), \\
    E_2^{0,2}=& \Hom\left(\Omega_{2}^\text{Spin}(\pt),2\pi \R/\Z\right)\cong \Z_2.
    \end{array}
\end{equation}
This description of the classification of 2d fermionic SPTs is essentially equivalent to the one in \cite{Gaiotto:2015zta,brumfiel2016pontrjagin,Wang:2017moj,Wang:2018pdc}.

Moreover, there is a coboundary map $H^2(BG,2\pi \R/\Z)\rightarrow H^3(BG,\Z)$ induced by the short exact sequence $\Z\rightarrow \R \rightarrow 2\pi \R/\Z$ of the coefficients. Altogether, this gives a relation between the choice of the generalized theta-angle $\alpha \in \Hom\left(\Omega_{2}^\text{Spin}(\pt),2\pi \R/\Z\right)$ and a choice of \textit{both} a spin and a string structure on sigma-model target $X$. At the intuitive level, the need of choice of spin and string structure on the target arises from the ambiguity of the definition of the Pfaffian arising after integrating fermions on the space of maps from the world-sheet to the target $X=Y/G$, whose topology is in part captured by $BG$. The different topological terms in the gauge theory thus change the dependence of the Pfaffian on the homotopy type of such map.

\subsection{A simple example of $\CN = (0,1)$ GLSM/sigma-model correspondence}
\label{sec:RPN-example}

As a simple illustration, consider $(0,1)$ GLSM with sufficiently large number $n_b=N$ (namely $N>4$) of scalar superfields, $n_f=1$ Fermi multiplet, the superpotential term $\Gamma\left(\sum_{i=1}^N(\phi^i)^2-r\right)$ with $r>0$ and a $G=\Z_2$ gauge field with respect to which the scalar superfields are charged and the Fermi superfield is neutral. In this case, the filtration (\ref{AH-filtration}) splits and
\begin{equation}
    \begin{array}{rl}
    E_2^{2,0}=& H^2(B\Z_2,2\pi \R/\Z)\cong 0, \\
    E_2^{1,1}=& H^1(B\Z_2,\Z_2)\cong \Z_2, \\
    E_2^{0,2}=& \Hom\left(\Omega_{2}^\text{Spin}(\pt),2\pi \R/\Z\right)\cong \Z_2.
    \end{array}
    \label{AHSS-BZ2}
\end{equation}
The choice of spin structure on a worldsheet $\Sigma$ can be encoded in the choice of a quadratic form
\begin{equation}
    q_\Sigma:\;H^1(\Sigma,\Z_2)\longrightarrow \Z_2
\end{equation}
satisfying
\begin{equation}
    q_\Sigma(a+b)=q_\Sigma(a)+q_\Sigma(b)+a\cup b.
\end{equation}
The corresponding topological term then can be explicitly written as
\begin{equation}
    S_\text{top}=\pi \left(\alpha_1q_\Sigma(f_G^*(x))+\alpha_2\text{Arf}(\Sigma)\right)
    \label{top-term-Z2}
\end{equation}
where $\alpha_i\in \Z_2$, $x$ is the generator of $H^1(B\Z_2,\Z_2)$, and $\text{Arf}(\Sigma)$ is the Arf invariant of the spin surface $\Sigma$ (equivalently, Arf invariant of the quadratic form $q_\Sigma$). This theory does not have a global gauge anomaly if the number of charged chiral fermions is a multiple of 8 \cite{gu2014effect,Kapustin:2014dxa}. This is related to the Fidkowski--Kitaev anomaly \cite{fidkowski2011topological} in one less dimension by the Smith isomorphism, i.e.~$N\equiv 0\pmod 8$, which corresponds to a vanishing of the corresponding element in $\Hom\left(\Omega^\text{Spin}_3(BG),2\pi \R/\Z\right)\cong \Z_8$. 

The space of solutions to the equations produced by the superpotential is a sphere
\begin{equation}
    Y=\left\{\sum_{i=1}^N(\phi^i)^2=r\right\} \cong S^{N-1}
\end{equation}
and has a unique spin and string structure (recall that $N>4$). Its $\Z_2$ quotient is the real projective space
\begin{equation}
    X=Y/\Z_2 \cong \mathbb{RP}^{N-1}
\end{equation}
and has the following cohomology with $\Z$ and $\Z_2$ coefficients:
\begin{equation}
    H^*(\mathbb{RP}^{N-1},\Z_2)=\Z_2[x]/x^{N},
\end{equation}
\begin{equation}
    H^p(\mathbb{RP}^{N-1},\Z)=
    \left\{
    \begin{array}{rl}
        \Z, & p=0, \\
        0, & p<N-1\text{ and $p$ odd}, \\
        \Z_2, & p<N-1\text{ and $p$ even}, \\
        \Z, & p=N-1\text{ odd, } \\
        \Z_2, & p=N-1\text{ even}.
    \end{array}
    \right.
\end{equation}
First, in order for the sigma-model to be well defined we need to require that $w_1(T\mathbb{RP}^{N-1})=0$ and $w_2(T\mathbb{RP}^{N-1})=0$. Using the following known expression for the total Stiefel--Whitney class of the tangent bundle
\begin{equation}
 w(T\mathbb{RP}^{N-1})=(1+x)^{N} = 1+Nx+\frac{N(N-1)}{2}x^2+\cdots \; \in \Z_2[x]/x^{N},
  \label{total-SW-RP}
\end{equation}
one can see that these conditions are equivalent to the statement that $N\equiv0\pmod 4$. Furthermore, we need to require that $p_1\left(T\mathbb{RP}^{N-1}\right)/2 =0$ in the integral cohomology. Given that $H^4(T\mathbb{RP}^{N-1})\cong \Z_2$, it is enough to require that 
\begin{equation}
    \frac12 p_1\left(T\mathbb{RP}^{N-1}\right) \bmod 2 = w_4(T\mathbb{RP}^{N-1}) =0 \quad \in \quad H^4(T\mathbb{RP}^{N-1},\Z_2) \,.
\end{equation}
Using again the explicit expression (\ref{total-SW-RP}), we arrive at the condition $N\equiv 0\pmod 8$. This is exactly the condition to have an anomaly-free $\Z_2$ gauge symmetry in the gauge theory description!

Since $H^3(\mathbb{RP}^{N-1},\Z)=0$ and $H^1(\mathbb{RP}^{N-1},\Z_2)=\Z_2$ there is a unique string structure and two different spin structures, which is in agreement with the classification (\ref{AHSS-BZ2}) of topological terms for the $\Z_2$ gauge field. Namely, different spin structures on $\mathbb{RP}^{N-1}$ correspond to different values of $\alpha_1$ in the topological part of the GLSM action (\ref{top-term-Z2}). The second term in (\ref{top-term-Z2}) is transferred directly to the sigma-model description. Note that in the GLSM, the sum over gauge bundles, that is the homotopy classes of the map $f_G:\Sigma\rightarrow B\Z_2$, can be interpreted as the sum of the homotopy classes of the maps $\phi: \Sigma\rightarrow X=\mathbb{RP}^{N-1}$ in the sigma-model description. One way to see this is to realize $B\Z_2=\mathbb{RP}^\infty$ and the sigma-model target as its subspace $X=\mathbb{RP}^{N-1}$. Then, taking into account that $N$ is sufficiently large, we can deform $f_G$ so that its image lies inside $\mathbb{RP}^{N-1}\subset \mathbb{RP}^\infty$. That is, $f_G$ and $\phi$ are related by the condition that the diagram 
\begin{equation}
\begin{tikzcd}
 & B\Z_2=\mathbb{RP}^\infty &  
\\
 \Sigma \ar[ur,"f_G"] \ar[r,"\phi"] & X=\mathbb{RP}^{N-1} \ar[u,hook] 
\end{tikzcd}
\end{equation}
commutes up to homotopy.

\subsection{Elliptic genus}
    \label{sec:elliptic-genus}

The elliptic genus (in the Ramond--Ramond sector) of a $(0,1)$ supersymmetric quantum field theory (not necessarily superconformal) can be defined as the index of the supercharge operator $Q_+$ acting on the Hilbert space $\CH$ of the theory on a circle, refined by the $U(1)$ rotation symmetry of the circle \cite{Witten:1986bf},
\begin{equation}
    I(q):=\Tr_{\CH|_{Q_+=0}}(-1)^F q^{P}.
\end{equation}
Here the momentum operator $P$ generate the $U(1)$ isometry of the circle and $F$ is the fermion parity. Note that instead of the total fermion parity $F$ in principle one can use only the right-moving fermion parity $F_R$. However, the latter in general can be explicitly broken, especially away from the superconformal point (e.g.~by the mass terms or Yukawa couplings). The same argument as for the Witten index in quantum mechanics  shows that such refined index should be invariant under continuous deformations of the theory, and also equal to the partition function of the theory on $T^2$ with complex structure $\tau$ (which is related to $q$ via $q:=e^{2\pi i \tau}$) and odd spin structure (i.e.~with periodic-periodic boundary conditions on the fermions). At the conformal point, it is also given by 
\begin{equation}
    I(q)= \Tr_{\CH}(-1)^F\,q^{L_0-c_L/24}\bar{q}^{\bar{L}_0-c_R/24},
\end{equation}
where $L_0$ and $\bar{L}_0$ are the standard generators of the Virasoro algebra and $L_0-\bar{L}_0=P$.

For the sigma-model, the elliptic genus can be computed as an integral of a certain characteristic class. Assume the target $X$ is even-dimensional and the bundle of left-moving fermions $E$ is of even rank. Then,
\begin{equation}
    I(q)=\int_{X} \prod_{i=1}^{\dim X/2}\frac{\xi_i}{\hat{\theta}(\xi_i;\tau)}
    \prod_{a=1}^{\rank E/2}\hat{\theta}(\eta_a;\tau)
    \label{elliptic-genus-integral}
\end{equation}
where $\pm\xi_i$ and $\pm\eta_a$ are Chern roots\footnote{Here and below, by Chern roots of a real rank $2n$ vector bundle with $O(2n)$ or $SO(2n)$ connection $\nabla$ we mean the entries $\pm\xi_i,i=1,\ldots,n$ appearing in the curvature 2-form brought by the adjoint action to the form
\begin{equation}
    F_\nabla=\left(
    \begin{array}{ccccc}
    0 & \xi_1 & 0 & 0 & \ldots \\
    -\xi_1 & 0 & 0 & 0 & \ldots \\
    0 & 0 & 0 & \xi_2 & \ldots \\
    0  & 0 & -\xi_2 & 0 & \ldots \\
    \ldots & \ldots & \ldots & \ldots & \ldots \\
    \end{array}
    \right).
\end{equation}
Alternatively, $\pm\xi_i,i=1,\ldots,n$ can be understood as the standard Chern roots of the complexified bundle. Note that the Weyl symmetry of $O(2n)$ includes the change of signs of any number of $\xi_i$'s while the Weyl symmetry of $SO(2n)$ only includes the change of sign of even number of change of signs of any number of $\xi_i$'s. That is why in the oriented case, there is a well-defined Euler class $\prod_{i=1}^n\xi_i$.
} of the vector bundles $TX$ and $E$, respectively, and\footnote{Up to a simple overall factor, $\hat{\theta}(u;\tau)$ coincides with the classical Jacobi theta function $\theta_1(u;\tau)$.}
\begin{equation}
    \hat{\theta}(u;\tau):=q^{1/12}(e^{u/2}-e^{-u/2})\prod_{n\geq 1}(1-q^ne^u)(1-q^ne^{-u}).
\end{equation}
It is also useful to introduce the normalized elliptic genus
\begin{equation}
    \Phi(q):=\eta(q)^{\dim X-\rank E}\,I(q) \label{index-renorm},
\end{equation}
where
\begin{equation}
\eta(q):=q^{1/24}\prod_{n\geq 1}(1-q^n)
\end{equation}
is the Dedekind eta function.

Note that in principle the expression (\ref{elliptic-genus-integral}) can be evaluated for any real manifold $X$ with a real vector bundle $E$. In general, however, $\Phi(q)\in \Q[[q]]$ may not have integer coefficients, and $\Phi(q)\in \Z[[q]]$ only when the anomaly cancellation condition $w_2(TX)-w_2(E)=0$ is satisfied. Moreover, if the second condition, $(p_1(TX)-p_1(E))/2=0$, is satisfied, $\Phi(q)$ is a modular form of weight $(\dim X-\rank E)/2$. The normalization factor in (\ref{index-renorm}) is chosen to trade the multiplier system for $SL(2,\Z)$ modular transformations --- that arises due to the gravitational anomaly $c_R-c_L=\frac{\dim X-\rank E}{2}$ --- for a non-trivial modular weight. At the same time, the value of $I(q)$ does not depend on a choice of the spin or string sctructure.

When a theory has flavor symmetry $H$, i.e.~a symmetry that commutes with the supercharge and the fermion number operator, it is possible to refine the elliptic genus by inserting the image of an element $h\in H$ of the flavor group under the representation map $\rho_H: H\rightarrow \mathrm{End}(\CH)$:
\begin{equation}
    I(h;q):=\Tr_{\CH|_{Q_+=0}}(-1)^F q^{P} \rho_H(h).
\end{equation}
Such refined elliptic genus is often referred to as \textit{equivariant} or \textit{flavored} elliptic genus. Of course, the function $I(h;q)$ only depends on the conjugacy class of $h$. The coefficients of the expansion of the elliptic genus in $q$ can be understood as the elements of the (real) representation ring $RO(H)$. The modular transformation properties of $I(h;q)$ are as of Jacobi-like modular form (the classical Jacobi forms correspond to the case $H=U(1)$) with the index determined by the 't Hooft anomaly of $H$.

In case $H$ is a connected simple Lie group, $h$ can be assumed to be an element of a maximal torus of $T_H\subset H$. Suppose the theory has a non-linear sigma-model description with $H$ realized as an isometry of the target space $X$, which lifts to an isometry of $E$ that commutes with the projection map $E\rightarrow X$ (i.e.~$E$ is an $H$-equivariant vector bundle over $X$). Then, $I(q;h)$ can still be computed using the formula (\ref{elliptic-genus-integral}), but equivariantly with respect to $T_H$. To do that, one can use the generalized Atiyah--Bott localization theorem. Namely, let $h_\ell$ be coordinates of $h$ on the maximal torus $T_H\cong (\R/2\pi \Z)^{\rank H}$ and $\alpha\in H^*_G(X,\C)$ an arbitrary class in equivariant cohomology of X.\footnote{As $H^*(X)\otimes \C[h_i] \rightarrow H^*_G(X)$ is surjective, we can always represent $\alpha$ by an element in $H^*(X)\otimes \C[h_i]$.} Then, we have \cite{atiyah1988moment,berline1983zeros}
\begin{equation}
    \int_X \alpha = \sum_{C\subset F}\int_C \cfrac{\alpha}{\text{Eu} (N_C)},
    \label{Atiyah-Bott}
\end{equation}
where the sum is performed over connected components $C$ of the fixed point set $F\subset X$ of $T_H$ action, $N_C$ is the bundle normal to $C$, and $\text{Eu}$ is the equivariant Euler class. Applying this general formula to (\ref{elliptic-genus-integral}) in the case when fixed points are isolated, $F=\{x_i\}$, we have
\begin{equation}
    I(h;q)=\sum_{i} \cfrac{ \prod\limits_{a=1}^{\rank E/2}\hat{\theta}\left(\sum\limits_{\ell=1}^{\rank H}\eta_{a,\ell}^{(i)}h_\ell;\tau\right)}{\prod\limits_{i=1}^{\dim X/2}\hat{\theta}\left(\sum\limits_{\ell=1}^{\rank H}\xi_{i,\ell}^{(i)}h_\ell;\tau\right)},
    \label{Atiyah-Bott-EG}
\end{equation}
where $\eta_{a,\ell}^{(i)}$ and $\xi_{i,\ell}^{(i)}$ are weights of the action of $T_H$ on the fibers $E_{x_i}$ and $T_{x_i}X$ at the fixed points respectively.

It is worth noting that the non-equivariant elliptic genus is identically zero when $\rank E>\dim X$. This is consistent with the fact that such sigma model can be continuously deformed (preserving $\mathcal{N}=(0,1)$ supersymmetry) to a theory that \textit{spontaneously} breaks supersymmetry. Such a deformation can be done by turning on a superpotential corresponding to a generic smooth section of the $E$. There will be no zero of
this section and thus supersymmetry will be spontaneously broken on the semiclassical level. This indicates that the refinement of the elliptic genus valued in the coefficient ring of TMF should also vanish identically for such models. This is however in general not true for the equivariant elliptic genus and, therefore, also its refinement valued in the coefficient ring of equivariant TMF. This is consistent with the fact that the latter is an invariant only under deformations preserving the corresponding flavor symmetries. In particular, one is only allowed to turn on a superpotential which is an $H$-equivariant section of $E$.


\section{A family of 2d $(0,1)$ gauge theories}
\label{sec:SO-GLSMs}

\subsection{UV Lagrangian description and gauge anomaly cancellation}

\begin{figure}[h]
\centering
\begin{tikzpicture}[every loop/.style={min distance=15mm},
roundnode/.style={circle, draw=black, very thick, minimum size=7mm},
squarednode/.style={rectangle, draw=black, very thick, minimum size=5mm},
]
\node[roundnode]     (gauge) at (0,0)   {$n_3$};
\node[squarednode]   (flavor1)    at (-3,2)  {$N_2$};
\node[squarednode]   (flavor2)    at (-3,-2)  {$N_1$};
\node[squarednode]   (flavor3)    at (3,0)  {$N_3$};


\draw[black, thick] (gauge)  -- (flavor1) node[midway,above]{$P$};
\draw[black, thick] (gauge) -- (flavor2) node[midway,above]{$\Phi$}; 
\draw[black, thick, dashed] (gauge) -- (flavor3) node[midway,above]{$\Psi$};
\draw[black, thick, dashed] (flavor1) -- (flavor2) node[midway,left]{$\Gamma$}; 
\path        (gauge)   edge[loop,dashed,black,thick,out=40,in=100,looseness=10] node[above]  {$\Sigma$} (gauge);

\end{tikzpicture}
\caption{The quiver diagram for our 2d $(0,1)$ SQCD with gauge group $SO(n_3)$. Each solid line represents a 2d $(0,1)$ scalar multiplet, whereas a dashed line represents a Fermi multiplet.}
\label{fig:quiver}
\end{figure}
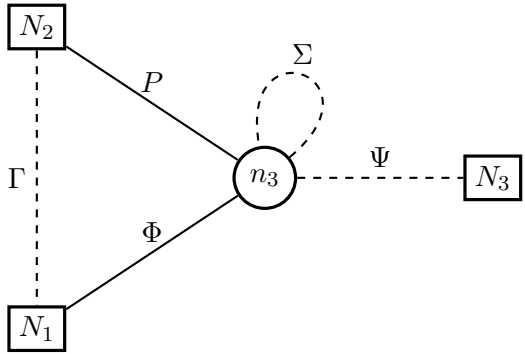

Consider a family of 2d $(0,1)$ gauge theories with $SO(n_3)$ gauge group and $SO(N_1)\times SO(N_2)\times SO(N_3)$ flavor symmetry, with the scalar and Fermi multiplets in the representations specified in the Table \ref{table:matter-content}. This matter content can also be summarized in a quiver diagram shown in Figure \ref{fig:quiver}.

\begin{table}[h]
    \centering
    \begin{tabular}{|c|c|c|c|c|c|c|}
    \hline
    type & symbol & $SO(n_3)$ & $SO(N_1)$ & $SO(N_2)$ & $SO(N_3)$ \\
    \hline
    \hline
    scalar    & $\Phi$ &  vector & vector & singlet & singlet\\
        \hline
    scalar    & $P$ &  vector & singlet  & vector & singlet \\
        \hline
    Fermi    & $\Psi$ &  vector & singlet  & singlet & vector         \\
        \hline
    Fermi     & $\Gamma$ &  singlet & vector  & vector & singlet \\
         \hline
    Fermi     & $\Sigma$ &  symmetric & singlet  & singlet & singlet \\
         \hline
    \end{tabular}
    \caption{The field content of our 2d $(0,1)$ SQCD with gauge group $SO(n_3)$.}
    \label{table:matter-content}
\end{table}
We also introduce the following superpotential terms,
\begin{equation}
    \int d^2xd\theta^+\;\sum_{\alpha,\beta=1}^{n_3}\Sigma^{\alpha\beta}\left(A\sum_{i=1}^{N_1}\Phi^{\alpha}_i\Phi^{\beta}_i+B\sum_{\ell=1}^{N_2}P^{\alpha}_\ell P^\beta_\ell-C\;\delta^{\alpha\beta}\right) +\sum_{i=1}^{N_1}\sum_{\ell=1}^{N_2}\sum_{\alpha=1}^{n_3}\Gamma_{i,\ell}\Phi^{\alpha}_iP^\alpha_\ell,
\end{equation}
where $A,B,C$ are some real constants. As will be explained later, the infrared dynamics of the theory depends only on their relative signs.

The possible generalized theta-angles in this class of theories are classified by
\begin{equation}
    \Hom\left(\Omega^\text{Spin}_2(BSO(n_3)),2\pi \R/\Z\right)\cong \Z_2^2
\end{equation}
so that the topological term reads
\begin{equation}
    S_\text{top}=\pi \left(\alpha_1 w_2(SO(n_3))+\alpha_2\text{Arf}(\Sigma)\right)
    \label{GLSM-SO-top-term}
\end{equation}
with $\alpha_i\in \Z_2$. In order for the theory to be well defined, one needs to show that the $SO(n_3)$ symmetry is not anomalous. At the perturbative level, as explained in Section \ref{sec:GLSM-review}, the anomaly cancellation condition is given in terms of indices of the representation. For convenience we present a table of indices and dimensions of relevant representations of $SO(n)$ in Table \ref{table:SOn-reps}.
\begin{table}[h]
\begin{center}
    \begin{tabular}{|c|c|c|c|}
    \hline
        $\rho$ &  vector & adjoint & symmetric\\
        \hline
        \hline
        $\dim \rho$ & $n$ & $\frac{n(n-1)}{2}$ & $\frac{n(n+1)}{2}$ \\
        \hline
        $T(\rho)$ & $1$ & $n-2$ & $n+2$ \\
        \hline
    \end{tabular}
\end{center}
    \caption{Dimension and indices of some basic $SO(n)$ representations.}
    \label{table:SOn-reps}
\end{table}
    
Note that the symmetric representation is reducible. It decomposes into traceless symmetric and a singlet. The singlet, of course, does not contribute to the anomaly. Therefore, we have the following condition on the anomaly cancellation in this class of 2d $(0,1)$ gauge theories:
\begin{equation}
    N_1+N_2-N_3-(n_3+2)-(n_3-2)=0 \qquad\Leftrightarrow \qquad n_3=\frac{N_1+N_2-N_3}{2}.
\end{equation}
In principle, one also has to make sure that there is no global gauge anomaly.\footnote{In the literature, sometime the phrase ``global anomaly'' is used to refer to the anomaly of a global symmetry. However, in this paper, the term ``global anomaly'' is reserved for the anomaly with respect to large gauge transformations.} In this particular case, the group classifying global anomalies, $\Hom\left(\Omega_3^\text{Spin}(BSO(n)),U(1)\right)$, is trivial. This, however, is no longer the case when the gauge group is replaced by $O(n)$, as will be discussed in Sections \ref{sec:On-gauge}. Similarly, there is also no possible mixed anomaly (pertubative or global) between $SO(N_i)$ flavor symmetries and the gauge symmetry. We will also confirm the canceallation of the anomalies in the sigma-model description. 

At the classical level, the theory actually has larger flavor symmetries, $O(N_i)$, under which the superfields $P,\Phi,\Gamma$, and $\Psi$ transform in vector representations. However, at the quantum level they generically have global mixed anomalies with the $SO(n_3)$ gauge group corresponding to the 3d SPTs with action $w_2(SO(n_3)) w_1(O(N_i))$. 

\subsection{Triality: Anomaly matching}

In the previous section we described a family of $(0,1)$ gauge theories labeled by ordered triples $(N_1,N_2,N_3)\in \Z^3_+$ and further parametrized by $(A,B,C)\in \R^3$. The gauge anomaly cancellation condition unambiguously determines the rank of the gauge group $SO(n_3)$ with $n_3=\frac{N_1+N_2-N_3}{2}$. This gives a restriction on the triples,
\begin{equation}
 N_1+N_2-N_3\in 2\Z_+.
\end{equation}

We conjecture that the theories labeled by permuted triples $(N_1,N_2,N_3)$ (assuming they all satisfy the condition above), when the parameters $(A,B,C)$ are in a certain cone in $\R^3$, all flow to the same infrared fixed point, which is a non-trivial $(0,1)$ SCFT. The dual theories have gauge groups $SO(n_i)$ where
\begin{equation}
    n_i=\cfrac{N_1+N_2+N_3}{2}-N_i,\qquad i=1,2,3.
\end{equation}
The proposed \textit{triality} is analogous to a triality of 2d SQCD models with larger $\CN=(0,2)$ supersymmetry \cite{Gadde:2013lxa}. 

The first simple check of the conjecture is the matching of 't Hooft anomalies of the theories that are supposed to be dual. The gravitational anomaly of a theory labeled by the triple $(N_1,N_2,N_3)$ is given by the difference between the numbers of left and right-moving fermions:
\begin{multline}
    2(c_R-c_L)=n_3N_1+n_3N_2-n_3N_3-N_1N_2-\frac{n_3(n_3-1)}{2}-
    \frac{n_3(n_3+1)}{2}=\\
    = \frac{1}{4}(N_1^2+N_2^2+N_3^2)-\frac{1}{2}(N_1N_2+N_2N_3+N_3N_1)
\end{multline}
which is indeed symmetric under permutations of $N_1,N_2,N_3$. The coefficients $k_i$ for the 't Hooft anomalies of $SO(N_i)$ flavor symmetries read
\begin{equation}
\begin{array}{ccc}
    k_1=&-n_3+N_2&=\frac{N_2+N_3-N_1}{2}, \\
    k_2=&-n_3+N_1&=\frac{N_3+N_1-N_2}{2}, \\
    k_3=&n_3&=\frac{N_1+N_2-N_3}{2},
    \label{flavor-anomaly-coefs}
\end{array}
\end{equation}
which are also symmetric under permutation of the indices $1,2,3$.

\subsection{Sigma-model description}
\label{sec:quiver-to-sigma}

For the gauge theory in the family considered above, the superpotential term involving Fermi multiplets $\Gamma$ gives the following equations on the bosonic components $\phi,p$ of the scalar superfields $\Phi,P$,
\begin{equation}
    \phi^\alpha_i p^\alpha_\ell =0,\qquad \forall\, i,\ell.
    \label{system-orthogonality}
\end{equation}
This system can be interpreted as the condition that the components $\phi_i^\alpha$ and $p^\alpha_\ell$ form two orthogonal collections of $N_1$ and $N_2$ vectors in $\R^{n_3}$ respectively.

The superpotential terms that involve the Fermi multiplet $\Sigma$ transforming in the symmetric representation of the gauge group $SO(n_3)$ gives the following equations, analogous to D-terms of higher-SUSY models:
\begin{equation}
    \left\{\begin{array}{ccl}
    A\sum_{i=1}^{N_1}(\phi^\alpha_i)^2+B\sum_{\ell=1}^{N_2}(p^\alpha_\ell)^2 & =C, & \forall \alpha, \\
    A\sum_{i=1}^{N_1}\phi^\alpha_i\phi^\beta_i
    +B\sum_{\ell=1}^{N_2}p^\alpha_\ell p^\beta_\ell & =0, &
    \forall  \alpha\neq \beta.
    \end{array}
    \right.
    \label{system-normalization}
\end{equation}
The topology of the space of solutions of the combined system of equations  depends drastically on the relative signs of $A,B$, and $C$. There are 4 essential cases (assuming neither of the constants vanishes):

\begin{enumerate}
    \item $C/A<0,\, C/B<0$. In this case the space of solutions is obviously empty, i.e.~$X=\emptyset$. The $(0,1)$ supersymmetry is expected to be broken dynamically. 
    \item $C/A>0,\, C/B<0$. First, one can argue that necessarily $p^\alpha_\ell=0,\forall\, \alpha,\ell$. Suppose that this is not true and some of the vectors $\Vec{p}_\ell\neq 0 \in \mathbb{R}^{n_3}$. By using the $SO(n_3)$ gauge symmetry one can always rotate it so that it is aligned with the first basis vector, i.e.~$p^1_\ell\neq 0$ while $p^\alpha_\ell=0,\;\alpha\neq 1$. The equation (\ref{system-orthogonality}) then implies that $\phi^1_i=0$ for all $i$, and the first equation of (\ref{system-normalization}) simplifies to
    \begin{equation}
        B (\Vec{p}_\ell)^2=C,
    \end{equation}
    which has no solutions due to $C/B<0$. Having shown that one necessarily has $p^\alpha_\ell=0,\forall \alpha,\ell$, the system of equations (\ref{system-normalization}) then simplifies to
\begin{equation}
    \left\{\begin{array}{ccl}
    A\sum_{i=1}^{N_1}(\phi^\alpha_i)^2 & =C, & \forall \alpha, \\
    A\sum_{i=1}^{N_1}\phi^\alpha_i\phi^\beta_i & =0, &
    \forall  \alpha\neq \beta.
    \end{array}
    \right.
\end{equation}
    which can be interpreted as the condition that the vectors $\Vec{\phi}^\alpha\in \R^{N_1},\;\alpha=1\ldots n_3$ are orthogonal and have squared norm $C/A$. The space of such collections of vectors, modulo $SO(n_3)$ rotations, is the \textit{oriented} real Grassmannian $\tilde{Gr}(n_3,N_1)$, i.e.~the space of oriented $n_3$-planes in $\R^{N_1}$. Therefore, the target space of the effective sigma model is
    \begin{equation}
        X=\tilde{Gr}(n_3,N_1)\cong \frac{SO(N_1)}{SO(n_3)\times SO(N_1-n_3)}.
    \end{equation}
    Following (\ref{bundle-from-GLSM}), the bundle of left-moving fermions is given by
    \begin{equation}
        E=S^{N_3}\oplus Q^{N_2}
    \end{equation}
    where $S$ is the rank-$n_3$ tautological bundle (i.e.~the bundle of the $n_3$-planes in the definition of the Grassmannian) and $S$ is the rank-$(N_1-n_3)$ orthogonal bundle, i.e.~the bundle of the $(N_1-n_3)$-planes orthogonal to the $n_3$-planes in the definition of the Grassmannian. The bundles $S^{N_3}$ and $Q^{N_2}$ arise as the massless modes of $\Gamma$ and $\Psi$ Fermi multiplets respectively. As explicitly shown in Appendix \ref{app:coh-grassmannian}, the anomaly cancellation conditions are indeed also satisfied in the sigma-model description.

    \item $C/A>0,\, C/B<0$. This case is essentially the same as the previous one, up to the exchange $N_1\leftrightarrow N_2$, $\Phi\leftrightarrow P$. Now one necessarily has $\phi^\alpha_i=0$ for all $ \alpha$ and $\ell$. The target space of the effective sigma-model is 
    \begin{equation}
        X=\tilde{Gr}(n_3,N_2),
    \end{equation} 
    and the bundle of left-moving fermions is 
    \begin{equation}
        E=S^{N_3}\oplus Q^{N_1}.
    \end{equation}
    
    \item $C/A>0,\, C/B>0$. This is the scenario when both $\phi$ and $p$ can be simultaneously non-zero. Suppose the vectors $\Vec{\phi}_i$ span a $k$-dimensional subspace in $\R^{n_3}$. From (\ref{system-orthogonality}) it follows that $\Vec{p}_\ell$ belongs to a $(n_3-k)$-dimensional orthogonal subspace. Using the $SO(n_3)$ gauge symmetry one can rotate this configuration so that $\phi_i^\alpha=0$ for $\alpha> k$ and $p_\ell^\alpha=0$ for $\alpha\leq k$. The equations (\ref{system-normalization}) then become 
    \begin{equation}
    \left\{\begin{array}{cll}
    \sum_{i=1}^{N_1}(\phi^\alpha_i)^2 & =C/A, & \forall \alpha\leq k, \\
    \sum_{i=1}^{N_1}\phi^\alpha_i\phi^\beta_i & =0, &
    \forall  \alpha\neq \beta \leq k, \\
\sum_{\ell=1}^{N_2}(p^\alpha_\ell)^2 & =C/B, & \forall \alpha>k, \\
    \sum_{\ell=1}^{N_2}p^\alpha_\ell p^\beta_\ell & =0, &
    \forall  \alpha\neq \beta >k.
    \end{array}
    \right.
\end{equation}
Taking into account the residual $S(O(k)\times O(n_3-k))$ subgroup of the $SO(n_3)$ gauge group which preserves such a splitting, the resulting moduli space of solutions is $\left(\tilde{Gr}(k,N_1)\times \tilde{Gr}(n_3-k,N_2)\right)/\Z_2$. Here $\Z_2$ acts as diagonal deck transformation of the forgetful cover $\tilde{Gr}(n,N)\rightarrow Gr(n,N)$, where $Gr(n,N)$ denote the Grassmannian of unoriented $n$-planes in $\R^N$. The target of the effective sigma model is then the disjoint union
\begin{equation}
    X=\bigsqcup_{k=0}^{n_3}\frac{\tilde{Gr}(k,N_1)\times \tilde{Gr}(n_3-k,N_2)}{\Z_2}
\end{equation}
Note that $\tilde{Gr}(0,N)=\tilde{Gr}(N,N)$ consist of two disjoint points.\footnote{For this reason, for self-consistency, one should define gauging $SO(0)$ group as taking two disjoint copies of the theory. At the same time, gauging $O(0)$ is a trivial operation.} The bundle of left-moving fermions over $k$-th connected component in the decomposition above is
\begin{equation}
    E_k=(S_1^{N_3}\oplus S_2^{N_3}\oplus (S_1\otimes
    S_2)\oplus (Q_1\otimes Q_2))/\Z_2
\end{equation}
where $S_1,Q_1$ and $S_2,Q_2$ are tautological and orthogonal bundles over $\tilde{Gr}(k,N_1)$ and $\tilde{Gr}(n_3-k,N_2)$ respectively.
\end{enumerate}

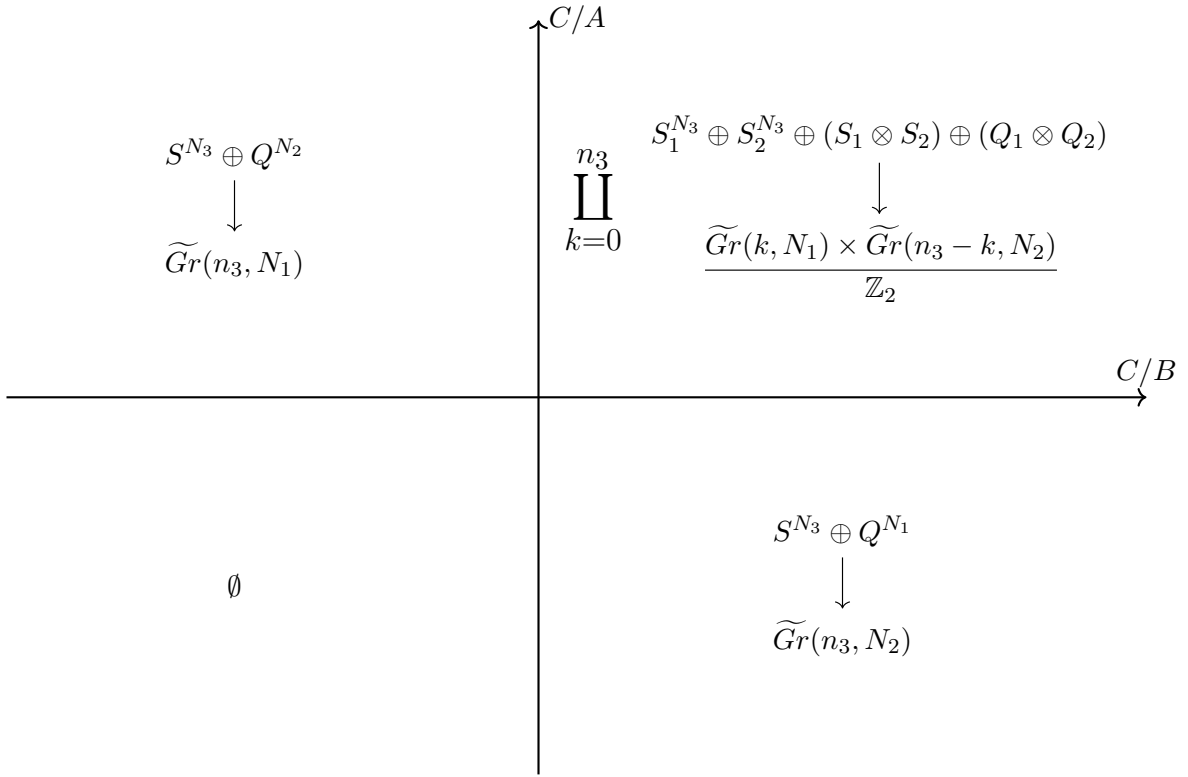
\begin{figure}[ht]
\centering
\begin{tikzpicture}
\draw[black, thick,->] (0,-5) -- (0,5);
\draw[black, thick,->] (-7,0) -- (8,0);
\node [right] at (0,5) {$C/A$};
\node [above] at (8,0) {$C/B$};
\node at (-4,-2.5) {$\emptyset$};
\node at (4,2.5) {$\text{\LARGE$\coprod\limits_{k=0}^{n_3}$}\;\; 
\begin{tikzcd}
 S_1^{N_3}\oplus S_2^{N_3}\oplus (S_1\otimes
    S_2)\oplus (Q_1\otimes Q_2) \ar[d]
\\
 \cfrac{\tilde{Gr}(k,N_1)\times \tilde{Gr}(n_3-k,N_2)}{\Z_2}
\end{tikzcd}
$};
\node at (-4,2.5) {$\begin{tikzcd}
 S^{N_3}\oplus Q^{N_2} \ar[d]
\\
 \tilde{Gr}(n_3,N_1)
\end{tikzcd}$};
\node at (4,-2.5) {$\begin{tikzcd}
 S^{N_3}\oplus Q^{N_1} \ar[d]
\\
 \tilde{Gr}(n_3,N_2)
\end{tikzcd}$};
\end{tikzpicture}
\caption{The classical phase diagram for the family of $SO(n_3)$ gauge theories. Globally, the space of couplings $(A,B,C)$ modulo rescalings is $\RP^2$.}
\label{fig:phases}
\end{figure}

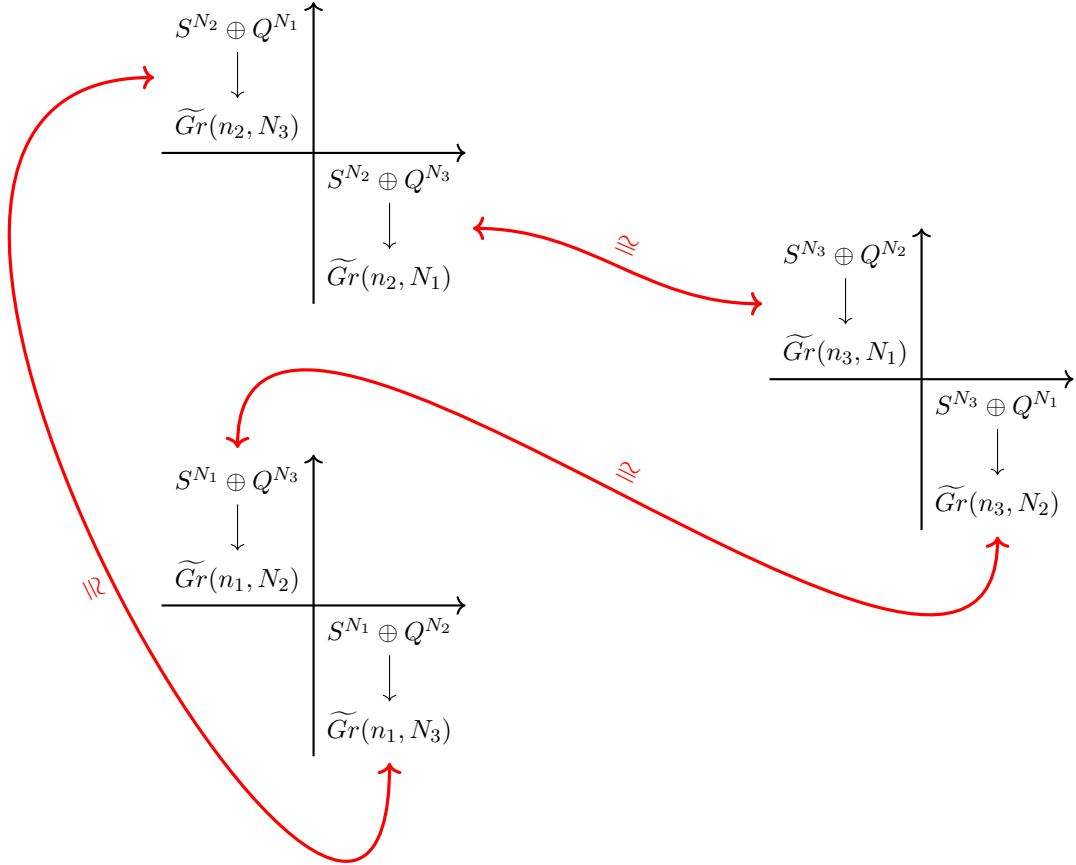
\begin{figure}[ht]
\centering
\begin{tikzpicture}
\draw[black, thick,->] (-4,1) -- (-4,5);
\draw[black, thick,->] (-6,3) -- (-2,3);
\node[scale=0.9] (T3NW) at (-5,4) {$\begin{tikzcd}
 S^{N_2}\oplus Q^{N_1} \ar[d]
\\
 \tilde{Gr}(n_2,N_3)
\end{tikzcd}$};
\node[scale=0.9] (T3SE) at (-3,2) {$\begin{tikzcd}
 S^{N_2}\oplus Q^{N_3} \ar[d]
\\
 \tilde{Gr}(n_2,N_1)
\end{tikzcd}$};
\draw[black, thick,->] (-4,-5) -- (-4,-1);
\draw[black, thick,->] (-6,-3) -- (-2,-3);
\node[scale=0.9] (T2NW) at (-5,-2) {$\begin{tikzcd}
 S^{N_1}\oplus Q^{N_3} \ar[d]
\\
 \tilde{Gr}(n_1,N_2)
\end{tikzcd}$};
\node[scale=0.9] (T2SE) at (-3,-4) {$\begin{tikzcd}
 S^{N_1}\oplus Q^{N_2} \ar[d]
\\
 \tilde{Gr}(n_1,N_3)
\end{tikzcd}$};
\draw[black, thick,->] (4,-2) -- (4,2);
\draw[black, thick,->] (2,0) -- (6,0);
\node[scale=0.9] (T1NW) at (3,1) {$\begin{tikzcd}
 S^{N_3}\oplus Q^{N_2} \ar[d]
\\
 \tilde{Gr}(n_3,N_1)
\end{tikzcd}$};
\node[scale=0.9] (T1SE) at (5,-1) {$\begin{tikzcd}
 S^{N_3}\oplus Q^{N_1} \ar[d]
\\
 \tilde{Gr}(n_3,N_2)
\end{tikzcd}$};
\draw[<->,very thick,red,overlay] (T3SE.east) to [out=0,in=180,edge node={node [sloped,above] {$\cong$}}](T1NW.west);
\draw[<->,very thick,red,overlay] (T3NW.west) to [out=180,in=-90,looseness=1.5,edge node={node [sloped,below] {$\cong$}}] (T2SE.south);
\draw[<->,very thick,red,overlay] (T2NW.north) to [out=90,in=-90,edge node={node [sloped,above] {$\cong$}}] (T1SE.south);
\node at (0,-6.5) {};
\node at (-7.5,0) {};
\end{tikzpicture}
\caption{Phase diagrams of the triple of GLSMs labelled by the permutations of $(N_1,N_2,N_3)$. The red arrows connect the regions where the theories are described by $(0,1)$ sigma-models with isomorphic targets and bundles of left-moving fermions.}
\label{fig:phases-triality}
\end{figure}

The phase structure of the theory is summarized in Figure \ref{fig:phases}. The UV limit corresponds to the targets being of large size, that is $C\rightarrow\infty$ (relative to $A$ and $B$). The exchange of the role of the $n_3$-dimensional subspace in $\R^{N_1}$ and the orthogonal $(N_1-n_3)$-dimensional subspace provides a canonical isomorphism between $\tilde{Gr}(n_3,N_1)$ and $\tilde{Gr}(N_1-n_3,N_1)\cong \tilde{Gr}(n_2,N_1)$, which exchanges the tautological and orthogonal bundles $S$ and $Q$. It follows that the targets and the bundles of left-moving fermions that appear in the lower right and upper left corners of the phase diagram for the theories that differ by the permutation of the triples $(N_1,N_2,N_3)$ are pairwise isomorphic, as illustrated in Figure \ref{fig:phases-triality}. It is however, not obvious if, for a GLSM with a given triple $(N_1,N_2,N_3)$, the theories in the lower left and upper right part of the phase diagram can be continuously deformed into each other. At the classical level, both regions touch the point where $C=0$, where the corresponding sigma-model targets become infinitesimally small and, therefore, quantum effects become strong. We conjecture that the theories from the two regions can be continuously deformed into each other at the quantum level (preserving $(0,1)$ supersymmetry). We will support this claim by matching equivariant elliptic genera in the next section.

Finally, let us comment on how to see the cancellation of the gauge anomalies from the sigma-model point of view. Consider, for example, the phase where the theory is described as a $(0,1)$ sigma-model with target $X=\tilde{Gr}(n_3,N_2)$ and the bundle $E=S^{N_3}\oplus Q^{N_1}$. First, one needs to check that $w_1(TX)=w_1(E)$ and $w_2(TX)=w_2(E)$. The mod-2 cohomology of the real Grassmannian and the explicit expressions for the Stiefel--Whitney classes of the tangent bundles is described in detail in Appendix \ref{app:coh-grassmannian}. Using the fact that $w(E)=(w(S))^{N_3}(w(Q))^{N_1}$, where $w$ denotes the total Stiefel--Whitney class, and the explicit formulas in the Appendix it is straightforward to check that those conditions are indeed satisfied. Checking $(p_1(TX)-p_1(E))/2=0$ in the \textit{integral} cohomology is a little more subtle. However, due to the fact that the torsion in the cohomology has only order-2 non-trivial elements, it is enough to check that $p_1(TX)=p_1(E)$ for \textit{rational} Pontryagin classes and also that
\begin{equation}
    \frac12\left(p_1(TX)-p_1(E)\right)\equiv 0 \pmod 2\quad\Leftrightarrow \quad
    w_4(TX)=w_4(E) \,.
\end{equation}
These conditions can be verified by using the explicit formulas for $w(TX)$ and rational $p_1(TX)$ in Appendix \ref{app:coh-grassmannian}.

The choice of the generalized spin and string structures (i.e.~the trivialization of $w_2(E)-w_2(TM)$ and $(p_1(TM)-p_1(E))/2$, respectively) in the sigma-model description is reflected as follows in the GLSM description. First, as follows from Appendix \ref{app:coh-grassmannian}, $H^1(X,\Z_2)=0$, so that there is a unique generalized spin structure. We also have $H^3(X,\Z)=\Z_2$, so there are just two different string structures.  They correspond to different values of $\alpha_1$ in the topological part of the gauge theory action (\ref{GLSM-SO-top-term}). The second term in (\ref{GLSM-SO-top-term}) is transferred directly to the sigma-model description.

\subsection{Triality: Elliptic genus}
\label{sec:triality-EG}

For a gauge theory with $(0,2)$ supersymmetry, there is a general formula for the equivariant elliptic genus that can be obtained by localization \cite{Gadde:2013ftv,Benini:2013xpa}. In the $(0,2)$ setting, the gauging has geometric interpretation of taking a K\"ahler quotient (unlike the ordinary quotient in the $(0,1)$ setting) and the formula can be interpreted as the Jeffrey--Kirwan localization theorem \cite{jeffrey1995localization,guillemin1996jeffrey} which relates integrals of characteristic classes on the quotient to the integrals of equivariant characteristic classes on the original space. The latter can be evaluated using Atiyah--Bott equivarint fixed-point localization theorem \cite{atiyah1988moment,berline1983zeros}.

For 2d $(0,1)$ theories, there is no available localization procedure, either from the path integral or from the geometric quotient point of view. However, given the non-linear sigma-model description of the theory, if the target space and the bundle have a non-trivial action of the flavor symmetry group, one can in principle still calculate the equivariant version of the integral (\ref{elliptic-genus-integral}) using the Atiyah--Bott localization theorem as was already mentioned in Section \ref{sec:elliptic-genus}.

Consider a GLSM from our family in the domain of superpotential couplings where it has NLSM description specified by the bundle 
\begin{equation}
    \begin{tikzcd}
 S^{N_3}\oplus Q^{N_2} \ar[d]
\\
 \tilde{Gr}(n_3,N_1)
\end{tikzcd}.
\end{equation}
There is an $SO(N_1)\times SO(N_2)\times SO(N_3)$ flavor-symmetry group acting on the bundle equivariantly (i.e.~the action commutes with the projection map). Note that only $SO(N_1)$ acts non-trivially on the base, while the groups $SO(N_2)$ and $SO(N_3)$ rotate the fibers at fixed points on the base. It is clear that the expression (\ref{Atiyah-Bott-EG}) vanishes unless $N_1,N_2,N_3$ and $n_3$ are all even (in particular dimensions of the fiber and the base should be both even), so we will assume that this is the case in the rest of this section. First, one needs to classify fixed points of the action of a maximal torus $T_{SO(N_1)}$ of $SO(N_1)$. For the details we refer to \cite{he2016localization}. Let us choose the maximal torus to be the subgroup $\prod_{i=1}^{N_1/2}SO(2)_{(i)}\subset SO(N_1)$ where $SO(2)_{(i)}$ rotates $\R^2_{(i)}\subset \R^{N_1}$ the 2-dimensional plane spanned by the $(2i-1)$-th and $(2i)$-th basis vectors in $\R^{N_1}$. The fixed points of the Grassmannian $\tilde{Gr}(n_3,N_1)$ will be configurations of $n_3$-planes inside $\R^{N_1}$ invariant under all such rotations. This happens when the $n_3$-plane, up to orientation, coincides with $\oplus_{i\in \SUB} \R^2_{(i)}$ where $\SUB\subset \{1,2,\ldots,N_1/2\}$ with length being $|\SUB|=n_3/2$. Taking into account two possible orientations we conclude that the fixed points are labeled by pairs: the subsets with length $n_3/2$ of the set of $N_1/2$ elements and an additional $\pm$ label,
\begin{equation}
    X\supset F=\{x_{(\SUB,\beta)}\}_{\SUB\in \{1,2,\ldots,N_1/2\},|\SUB|=n_3/2;\beta=\pm}.
\end{equation}

From this description of the set of fixed points, it is easy to see how the maximal torus of $SO(N_1)$ acts on the tautological and orthogonal bundles. Denote the equivariant parameters for $SO(2)_{(i)}$ subgroups defined above by $\lambda_i$, i.e.~$H^*_{SO(2)_{(i)}}(\pt)\cong \C[\lambda_i]$. Then, the equivariant Chern roots of the tautological bundle $S$ at a point $x_{\SUB,\beta}$ are $\{\pm \lambda_i\}_{i\in \SUB}$. The collection of the Chern roots is the same for both orientations of the $n_3$-plane $\beta=\pm$, since the change of the orientation corresponds to changing the sign of one of the roots. In particular, the equivariant Pontryagin classes, which are basic symmetric polynomials in $\lambda_i^2$ are the same for $\beta=\pm$. However, the equivariant Euler class of $S$ at $x_{\SUB,\pm}$ is $\pm \prod_{i\in \SUB}\lambda_i$.

Similarly, for the orthogonal bundle $Q$, the equivariant Chern roots at $x_{(\SUB,\beta)}$ are $\{\pm \lambda_i\}_{i\in \bar{\SUB}}$ where $\bar{\SUB}=\{1,2,\ldots,N_1/2\}\setminus \SUB$ is the complement of $\SUB$ with length $|\bar\SUB|=N_1-n_3$. And the equivariant Euler class is $\pm \prod_{i\in \bar\SUB}\lambda_i$ for $\beta=\pm$. Note that the deck transformation of the 2-fold cover $\tilde{Gr}(n_3,N_1)\rightarrow Gr(n_3,N_1)$ of the Grassmannian of unoriented planes will swap the two points $x_{\SUB,\pm}$. Such deck transformation flips the sign of one of the $\lambda_i,i\in \SUB$ and also one of $\lambda_j,j\in \bar\SUB$. This corresponds to the action of an element of the Weyl group of $SO(N_1)$ which is \textit{not} in the Weyl group of $SO(n_3)\times SO(N_1-n_3)$.

Now the Chern roots of $TX$ can be determined from the canonical isomorphism $TX\cong S\otimes Q$. So, its equivariant Chern roots are $\{\pm \lambda_i\pm \lambda_j\}_{i\in \SUB,j\in \bar{\SUB}}$. The Euler class is $\prod_{i\in \SUB,j\in \bar{\SUB}}(\lambda_j^2-\lambda_i^2)$ for both $\beta=\pm$. While the Euler class of $S$ and $Q$ both flip sign upon going from $x_{\SUB,+}$ to $x_{\SUB,-}$, the Euler class of their tensor product remains the same.  

Having understood the action of $SO(N_1)$ on the base and on the bundles $TX,\,S$ and $Q$ it remains to consider the action of $SO(N_2)$ and $SO(N_3)$. They act by rotating the fibers of $E=S^{N_3}\oplus Q^{N_1}$ in a canonical way. Denote by $\mu_r$ and $\nu_a$ the equivariant parameters for the maximal tori in $SO(N_2)$ and $SO(N_3)$ respectively, with $r=1\ldots N_2/2$ and $a=1\ldots N_3/2$. The equivariant Chern roots of $S^{N_3}$ at the fixed point $x_{\SUB,\beta}$ are then $\{\pm \lambda_i\pm \nu_a\}_{i\in \SUB,a=1\ldots N_3/2}$ and the Euler class is $\prod_{i\in \SUB,a=1\ldots N_3/2}(\lambda_i^2-\nu_a^2)$ for both $\beta=\pm$. Similarly, the equivariant Chern roots of $Q^{N_2}$ at the fixed point $x_{\SUB,\beta}$ are $\{\pm \lambda_i\pm \mu_r\}_{i\in \bar\SUB,r=1\ldots N_2/2}$ and the Euler class is $\prod_{i\in \bar\SUB,r=1\ldots N_2/2}(\lambda_i^2-\mu_r^2)$ for both $\beta=\pm$. 

Now we are ready to apply the Atiyah--Bott localization formula (\ref{Atiyah-Bott}) to the equivariant integral (\ref{elliptic-genus-integral}) and obtain the following explicit expression for the equivariant elliptic genus,
\begin{multline}
    I_{N_1,N_2,N_3}(\lambda,\mu,\nu;q)=\\
    =2\hspace{-2em}\sum_{\text{$
    \begin{array}{c}
\SUB\subset \{1,2,\ldots,N_1/2\} \\
|\SUB|=n_3/2
\end{array}$}
}\hspace{-2em}
    \cfrac{\prod\limits_{i\in\SUB}\prod\limits_{a=1}^{N_3/2}\hat\theta(\lambda_i-\nu_a;q)\hat\theta(\lambda_i+\nu_a;q)\prod\limits_{j\in\bar\SUB}\prod\limits_{r=1}^{N_2/2}\hat\theta(\lambda_j-\mu_r;q)\hat\theta(\lambda_j+\mu_r;q)}{\prod\limits_{i\in\SUB}\prod\limits_{j\in\bar\SUB}\hat\theta(\lambda_j-\lambda_i;q)\hat\theta(\lambda_j+\lambda_i;q)}.
    \label{elliptic-genus-result}
\end{multline}
The factor of $2$ is the result of the sum over the index $\beta=\pm$ in the overall sum over the fixed points $x_{(\SUB,\beta)}\in F\subset X$. Since the summand depends only on $\lambda_i^2$ it gives the same contribution to both choices.

As expected, the expression above for the elliptic genus is explicitly symmetric under the exchange $N_3\leftrightarrow N_2$, $\nu\leftrightarrow \mu$, up to a possible sign,
\begin{equation}
    I_{N_1,N_2,N_3}(\lambda,\mu,\nu;q)=\pm I_{N_1,N_3,N_2}(\lambda,\nu,\mu;q).
\end{equation}
This symmetry corresponds to the isomorphism $\tilde{Gr}(n_3,N_1)\cong \tilde{Gr}(N_1-n_3,N_1)= \tilde{Gr}(n_2,N_1)$ accompanied by the exchange of $S$ and $Q$ bundles, as indicated by one of the arrows in Figure \ref{fig:phases-triality}. A possible extra sign corresponds to a possible flip of the overall orientation. In this work we want to avoid subtleties related to the choice of overall orientation of the target and bundles and thus want to consider the theories that differ by the orientation to be equivalent. At the level of the Hilbert space, this equivalence is realized by shifting the fermion number of the vacuum by 1. Note that, in particular, in Section \ref{sec:quiver-to-sigma} we considered gauge theories that differ by a change of the signs of the superpotential couplings $(A,B,C)\rightarrow (-A,-B,-C)$ to be equivalent, but, to get back to the original action, one needs to accompany this with $\Sigma\rightarrow -\Sigma$, which (depending on $n_3$) could change the sign of the path integral. 

What is non-trivial is that (\ref{elliptic-genus-result}) is also invariant (again, up to a possible sign) under the exchange $N_1\leftrightarrow  N_2$, $\lambda \leftrightarrow \mu$,
\begin{equation}
    I_{N_1,N_2,N_3}(\lambda,\mu,\nu;q)=\pm I_{N_2,N_1,N_3}(\mu,\lambda,\nu;q)
\end{equation}
We have verified this property by comparing the first terms of the expansions of $I(\lambda,\mu,\nu;q)$ and $I(\mu,\lambda,\nu;q)$ in $q$ for multiple values of $N_1,\,N_2$ and $N_3$. This identity strongly supports the claim made in the previous section, namely that the theories in the lower right and upper left domains of the phase diagrams shown in Figures \ref{fig:phases} and \ref{fig:phases-triality} can be continuously deformed into each other at the quantum level.

Let us note that the elliptic genus vanishes if one sets equivariant parameters $\lambda,\mu,\nu$ to zero. As was discussed in a general setting  at the end of Section \ref{sec:elliptic-genus}, this is in accordance with the fact that an NLSM with $\rank E > \dim X$ can be continuously deformed to a theory with spontaneously broken supersymmetry if one forgets about flavor symmetries. On the level of GLSM a deformation to a theory with spontaneously broken supersymmetry can be done by turning on a constant term in the (0,1) superpotential associated to the Fermi multiplet $\Gamma$.

\section{Generalizations}
\label{sec:generalizations}

\subsection{$O(n)$ gauge group}
\label{sec:On-gauge}

In this section, we briefly comment on how the above analysis can be generalized to the case when $SO(n_3)$ gauge groups in the family of theories considered in Section \ref{sec:SO-GLSMs} is replaced by $O(n_3)$ gauge group. Namely, let us consider the family of theories with the same field content as in Table \ref{table:matter-content}, also depicted in the quiver diagram of Figure \ref{fig:quiver}. As we will see in a moment, the matter content will have to be slightly modified to ensure the cancellation of gauge anomalies.  

The perturbative anomaly is the same as in the $SO(n_3)$ case and remains canceled as long as
\begin{equation}
    n_3=\frac{N_1+N_2-N_3}{2}
\end{equation}
However, unlike the $SO(n_3)$ gauge theory, in principle there can be a global anomaly corresponding to non-trivial elements of $\Hom\left(\Omega_3^{\text{Spin}}(BO(n_3)),2\pi \R/\Z\right)$. In particular, the theory can have a $\Z_2$ global anomaly corresponding to the 3d SPT with the action $\pi\, w_2(O(n_3))w_1(O(n_3))$ and a $\Z_8$ global anomaly corresponding to the 3d topological term $\frac{\pi}{4}\text{ABK}[\text{PD}(w_1(O(n_3))]$, where $\text{ABK}[\text{PD}(a)]$ denotes the Arf--Brown--Kervaire invariant of a pin$^-$ surface representing the Poincar\'e dual to the class $a$ in the first mod-2 cohomology group. The latter anomaly is essentially the same as the anomaly in fermionic 2d theories with $\Z_2$ symmetry, which has already played an important role in the example considered in Section \ref{sec:RPN-example}. The $\Z_2$ symmetry here is any $\Z_2\subset O(n_3)$ generated by an element with determinant $-1$. The Majorana--Weyl fermions charged with respect to this $\Z_2$ contribute $\pm 1$ to this mod-8 anomaly depending on the chirality. Counting the total contribution to this mod-8 anomaly for our family of theories gives
\begin{equation}
    N_1+N_2-N_3-(n_3-1)-(n_3-1) \equiv 2 \pmod 8
\end{equation}
which is always non-zero. The easiest way to fix this is to add two real left-moving fermions $\Omega^i,\, i=1,2$ transforming in the determinant representation of $O(n_3)$. Such fermions are analogs of the $(0,2)$ Fermi multiplets transforming in the determinant representation of $U(n_3)$ gauge group of $(0,2)$ gauge theories participating in the $(0,2)$ triality of \cite{Gadde:2013lxa}. There, they were required to cancel the anomaly for the diagonal $U(1)$ of $U(N)$. 

We would like to argue that once this modification to the theory is made, all other possible global anomalies are also canceled. However, instead of considering cancellation of the anomalies in the GLSM setting, we will do so in the non-linear sigma-model description. The semi-classical analysis of Section \ref{sec:quiver-to-sigma} goes through with the oriented Grassmannian $\tilde{Gr}(n,N)$ replaced everywhere with the Grassmannian $Gr(n,N)$ of \textit{unoriented} $n$-planes in $\R^N$, and two copies of the determinant of the tautological bundle added to the bundles of left-moving fermions. For example, in the regime $A/C>0,\,B/C<0$ the non-linear sigma-model is now described in term of the bundle
\begin{equation}
    \begin{tikzcd}
 E=S^{N_3}\oplus Q^{N_2} \oplus (\det S)^{2}\ar[d]
\\
 X={Gr}(n_3,N_1)
\end{tikzcd}.
\end{equation}

Similar to the oriented case, the non-trivial torsion elements of $H^*(Gr(n,N))$ are all of order 2. So to ensure the cancellation of global anomalies (assuming the perturbative anomalies are already canceled) it is enough to require that
\begin{equation}
    \begin{array}{rl}
     w_1(TX) =& w_1(E), \\
     w_2(TX) =& w_2(E), \\
     w_4(TX) =& w_4(E).
    \end{array}
\end{equation}
Using the explicit formulas in Appendix \ref{app:coh-grassmannian}, one can check that these conditions are indeed satisfied. If $n_3$ and $N_1$ are sufficiently large, we have $H^1(X,\Z_2)\cong \Z_2$ and $H^3(X,\Z)\cong \Z_2$. Therefore there are two possible spin and string structures. The GLSM description can have the following topological terms (cf.~(\ref{top-term-Z2}) and (\ref{GLSM-SO-top-term})):
\begin{equation}
    S_\text{top}=\pi \left(\alpha_1q_\Sigma\left(w_1(O(n_3))\right)+\alpha_2\int_\Sigma w_2(O(n_3))+\alpha_3\text{Arf}(\Sigma)\right).
    \label{top-term-O}
\end{equation}
The different values of $\alpha_1\in \Z_2$ and $\alpha_2\in \Z_2$ correspond to different choices of the generalized spin and string structures. The last term in (\ref{top-term-O}) is directly transferred to the sigma-model description.

The calculation of the elliptic genus is also completely parallel to the one in Section \ref{sec:triality-EG} for the $SO$ case. Since now the target $X$ is the Grassmannian of \textit{unoriented} $n_3$-planes in $\mathbb{R}^{N_1}$, there will be half as many fixed points as for the $SO(N_1)$ action. Namely, the fixed points now are labeled just by subsets $\SUB \in\{1,2,\ldots,N_1/2\}$ of length $|\SUB|=n_3/2$. The orientation label $\beta=\pm$ should be forgotten. Because of the addition of $(\det S)^2$ to the bundle of left-moving fermions $E$, to get a non-zero result one should calculate the integral equivariantly with respect to the $SO(2)$ symmetry rotating $(\det S)^2=\det S\otimes \R^2$. Let us denote the corresponding equivariant parameter by $\zeta$. Physically, $e^{2\pi i \zeta}$ is the fugacity of $SO(2)$ flavor symmetry rotating the fermions $\Omega^i$ that transform in the determinant representation of $O(n_3)$ gauge group. The explicit expression for the elliptic genus then reads
\begin{multline}
    I_{N_1,N_2,N_3}^{O}(\lambda,\mu,\nu;\zeta;q)=\\
   \hat\theta(\zeta;q) \sum_{\text{$
    \begin{array}{c}
\SUB\subset \{1,2,\ldots,N_1/2\} \\
|\SUB|=n_3/2
\end{array}$}
}\hspace{-2em}
    \cfrac{\prod\limits_{i\in\SUB}\prod\limits_{a=1}^{N_3/2}\hat\theta(\lambda_i-\nu_a;q)\hat\theta(\lambda_i+\nu_a;q)\prod\limits_{j\in\bar\SUB}\prod\limits_{r=1}^{N_2/2}\hat\theta(\lambda_j-\mu_r;q)\hat\theta(\lambda_j+\mu_r;q)}{\prod\limits_{i\in\SUB}\prod\limits_{j\in\bar\SUB}\hat\theta(\lambda_j-\lambda_i;q)\hat\theta(\lambda_j+\lambda_i;q)}.
    \label{elliptic-genus-result-O}
\end{multline}
The result differs from (\ref{elliptic-genus-result}) by an overall factor independent of $\mu,\lambda,\nu$. Therefore, we still have a symmetry with respect to any permutation of $N_1,N_2,N_3$ accompanied by a corresponding permutation of $\mu,\lambda,\nu$.

\subsection{$Sp(2n)$ gauge group}
We conjecture that the analogous triality holds for a family of $(0,1)$ gauge theories with $Sp(2n)$ gauge groups\footnote{One way to define $Sp(2n)$ is as the subgroup of $GL(n,\mathbb{H})$ (linear transformations of $n$-dimensional quaternionic vector space) that preserves the quaternion-hermitian form
\begin{equation}
    \langle x,y\rangle = x_1\bar{y}_1+\ldots+x_n\bar{y}_n
\end{equation}
where bar denotes quaternionic conjugate.} obtained from the theories considered in Section \ref{sec:SO-GLSMs} via the known \textit{negative dimension} correspondence
\begin{equation}
    Sp(2n)\longleftrightarrow SO(-2n)
    \label{Sp-SO}
\end{equation}
with representations related according to the exchange
\begin{equation}
    \text{symmetrization} \longleftrightarrow \text{anti-symmetrization}.
\end{equation}

Namely, we want to consider a family of $(0,1)$ gauge theories with $Sp(n_3)$ gauge group and $Sp(N_1)\times Sp(N_2)\times Sp(N_3)$ flavor symmetry, where $n_3,N_1,N_2,N_3$ are all even. The matter content is specified in Table \ref{table:matter-content-Sp}.

\begin{table}[h]
    \centering
    \begin{tabular}{|c|c|c|c|c|c|c|}
    \hline
    type & symbol & $Sp(n_3)$ & $Sp(N_1)$ & $Sp(N_2)$ & $Sp(N_3)$ \\
    \hline
    \hline
    scalar    & $\Phi$ &  vector & vector & singlet & singlet\\
        \hline
    scalar    & $P$ &  vector & singlet  & vector & singlet \\
        \hline
    Fermi    & $\Psi$ &  vector & singlet  & singlet & vector         \\
        \hline
    Fermi     & $\Gamma$ &  singlet & vector  & vector & singlet \\
         \hline
    Fermi     & $\Sigma$ &  antisymmetric & singlet  & singlet & singlet \\
         \hline
    \end{tabular}
    \caption{The matter content of the family of $Sp(n_3)$ gauge theories.}
    \label{table:matter-content-Sp}
\end{table}
One can explicitly check that gauge anomaly is canceled, which also automatically  follows  from the correspondence (\ref{Sp-SO}).

The superpotential can be obtained from the one in the $SO(n)$ case by doing the formal replacements $n_3\rightarrow -n_3$, $N_i\rightarrow -N_i$. Instead of writing down the superpotential explicitly, it is more convenient to write down the corresponding constraints on the scalar fields, which contain the same amount of information. For this purpose, it is convenient to represent the fields $\phi$ and $p$, respectively, as $N_2/2\times n_3/2$ and $N_1/2\times n_3/2$ quaternion-valued fields. Then, the equations produced by the superpotential terms containing $\Gamma$ read
\begin{equation}
    \sum_{\alpha=1}^{n_3/2}\phi_{i}^\alpha \bar{p}^\alpha_\ell=0,\qquad \text{ for all $i=1,\ldots,N_1/2$ and $\ell=1,\ldots,N_2/2,$}
\end{equation}
where each constraint is viewed as an $\mathbb{H}$-valued equation. So in total there are $4(N_1/2\times N_2/2)=N_1N_2$ real equations, which is indeed the same as the number of real component of $\Gamma$. The equations produced by the superpotential terms containing $\Sigma$ read
\begin{equation}
    \left\{\begin{array}{ccl}
    A\sum_{i=1}^{N_1/2}||\phi^\alpha_i||^2+B\sum_{\ell=1}^{N_2/2}||p^\alpha_\ell||^2 & =C, & \forall \alpha=1,\ldots,n_3/2, \\
    A\sum_{i=1}^{N_1/2}\phi^\alpha_i\bar{\phi}^\beta_i
    +B\sum_{\ell=1}^{N_2/2}p^\alpha_\ell \bar{p}^\beta_\ell & =0, &
    \forall  \alpha\neq \beta.
    \end{array}
    \right.
\end{equation}
There are $n_3/2$ real equations and $n_3/2\times(n_3/2-1)/2$ quaternionic equations. In total, there are $n_3/2+4n_3/2\times(n_3/2-1)/2=n_3(n_3-1)/2$ real equations, which is precisely the number of real component of $\Sigma$. The analysis similar to the one done in the Section \ref{sec:quiver-to-sigma} then leads to the same NLSM descriptions, but with real Grassmannians replaced by quternionic ones. For example, in the regime where $A/C>0$ and $B/C<0$, the target space is
\begin{equation}
    X=HGr(n_3/2,N_1/2)\cong \frac{Sp(N_1)}{Sp(n_3)\times Sp(N_1-n_3)}
\end{equation}
where $HGr(n_3/2,N_1/2)$ is the Grassmannian of quaternionic $n_3/2$-planes in  $\mathbb{H}^{N_1/2}$.

Finally, doing the formal replacements $n_3\rightarrow -n_3$ and  $N_i\rightarrow -N_i$ in (\ref{elliptic-genus-result}) tells us that the expression for the elliptic genus will be the same as in $SO$ case.

\subsection{Quiver mutations}
As in the case of 2d $(0,2)$ trialities \cite{Gadde:2013lxa}, one can use the basic theories considered in Section \ref{sec:SO-GLSMs} to construct quiver theories with gauge groups of the form $SO(m_1)\times SO(m_2)\times \ldots$ by ``gluing'' the basic theories together. Namely, one can glue two basic theories by identifying one of the $SO(N_i)$ flavor symmetries of one theory with the $SO(n_3')$ gauge symmetry of the other, and vice versa. The absence of mixed anomaly ensures that this is a well-defined operation at the quantum level. 
This process can be continued by gluing the result with another basic $(0,1)$ SQCD and so on. 

By replacing any of the basic theories in the composed quiver, one gets a dual theory described by a ``mutated'' quiver. Such quiver mutations produce a network of infrared dualities between different quiver theories.

\subsection{Non-compact models}
\label{sec:appetizer}

So far, we mainly focused our attention on the IR physics of ``compact'' $(0,1)$ models, where all scalar fields are constrained to be in a compact region of the field space. For example, at the intermediate energy scales, our models flow through non-linear sigma-models with Grassmannian target manifolds $X$.

Relaxing this condition leads to another natural generalization of our SQCD-type models, where some of the scalars classically can take arbitrarily large values. A simple family of such ``non-compact'' $(0,1)$ models is obtained by taking trivial superpotential, $J=0$. Consider, for example, a 2d $(0,1)$ gauge theory with gauge group $G=SU(N_c)$ and scalar matter fields that furnish $N_f$ copies of the fundamental representation $V = \C^{N_c} \cong \R^{2N_c}$. A more general class of models can be obtained by incorporating $(0,1)$ Fermi multiplets in various representations of $G$, but, for simplicity, here we will only consider SQCD-type theories with scalar matter multiplets.

Then, in the notations of Section~\ref{sec:GLSM-review}, such non-compact models have $n_f=0$, $n_b = 2 N_c N_f$, and can be represented by a quiver diagram analogous to the one shown in Figure~\ref{fig:quiver},
\begin{equation}
\begin{tikzpicture}[every loop/.style={min distance=15mm},
roundnode/.style={circle, draw=black, very thick, minimum size=7mm},
squarednode/.style={rectangle, draw=black, very thick, minimum size=5mm},
]
\node[roundnode]     (gauge) at (0,0)   {$N_c$};
\node[squarednode]   (flavor3)    at (3,0)  {$N_f$};
\draw[black, thick] (gauge.east) -- (flavor3.west) node[midway,above]{$~$};
\end{tikzpicture}
\label{NcNfquiver}
\end{equation}
except that now one has to keep in mind that gauge node represented by a circle stands for the $SU$ rather than $SO$ gauge group. In a gauge theory with $SU(N_c)$ gauge group, the contribution of 2d $(0,1)$ vector multiplet to the perturbative gauge anomaly is $- N_c$. And, each ``flavor,'' equivalent to 2d $(0,2)$ chiral multiplet in the fundamental representation of $G$, contributes $+1$ to the perturbative gauge anomaly, cf. \cite{Gadde:2013lxa,Dedushenko:2017osi}. Therefore, the anomaly cancellation condition \eqref{TTTanom} in this case reads
\begin{equation}
N_f \; = \; N_c
\end{equation}
and the gravitational anomaly \eqref{gravanom} is
\begin{equation}
2(c_R-c_L) \; = \; (2N_f - N_c) N_c + 1
\end{equation}
There is no global, non-perturbative anomaly since $\Hom\left(\text{Tor}\,\Omega_3^\text{Spin} \big(BSU(N_c)\big),U(1)\right)$ is trivial.

In the notations of Section~\ref{sec:RGflow}, this family of non-compact $(0,1)$ models had $Y = \R^{2 N_c^2}$ and, therefore, at the intermediate energy scales flows to a NLSM with target space \eqref{XYGtarget}:
\begin{equation}
X \; = \; Y/G \; = \; \R^{2 N_c^2} / SU(N_c)
\label{noncomptarget}
\end{equation}
and a skyscraper sheaf (comprised from the fermions of the vector multiplet) supported at the origin of $X$. The quotient space $X$ is a cone on $S^{2 N_c^2 - 1} / SU(N_c)$, where $\R^{2 N_c^2}$ itself is viewed as a cone on $S^{2 N_c^2 - 1}$.

In order to understand the geometry of this quotient better, let us consider the simplest non-trivial theory in this family, namely the case of $N_c = N_f = 2$. At the intermediate energy scales it can be described by a NLSM with target space $X$ which is quotient of $Y = \R^{8} \cong \C^{4}$ by $G=SU(2)$. Note, $\C^4$ is a fundamental representation of the symmetry group $SU(4)$ and there are three homomorphisms $SU(2)_c \times SU(2)_f \to SU(4)$, which correspond to three different topological twists of 4d $\CN=4$ super-Yang-Mills \cite{Vafa:1994tf}. The one relevant for us here, in which $\C^4$ is realized as $N_f = 2$ copies of the fundamental representation of $SU(2)$, corresponds to the Hopf fibration $S^3 \to S^7 \to S^4$. In other words, the quotient space $S^7 / SU(2) \cong S^4$ and $X \cong \text{Cone} (S^4) = \R^5$. It is therefore tempting to conjecture that 2d $(0,1)$ SQCD with $N_c = N_f = 2$ in the IR flows to a theory of 5 free $(0,1)$ scalar multiplets.\footnote{Somewhat similar behavior was proposed in two-dimensional theories with larger supersymmetry~\cite{Aharony:2016jki}, where it was argued that RG flow flattens a conical target space into a free IR theory.}

A natural way to test this duality might be to start with 2d $\CN=(0,2)$ SQCD that has $N_c = 2$, $N_f = 4$ and is dual to 2d $\CN=(0,2)$ Landau-Ginzburg model with 6 chiral $(0,2)$ multiplets~\cite{Dedushenko:2017osi}. Then, just like in 3d dualities related by soft supersymmetry breaking \cite{Kachru:2016rui,Kachru:2016aon}, one would expect that deforming both sides of 2d $(0,2)$ duality \cite{Dedushenko:2017osi} by dual relevant operators that preserve only 2d $\CN=(0,1)$ supersymmetry one should end up with a 2d $(0,1)$ duality involving $N_c = N_f = 2$ SQCD considered here.

Another way to prove or disprove whether 2d $(0,1)$ SQCD with $N_c = N_f = 2$ flows to a theory of 5 free $(0,1)$ scalar multiplets is by realizing this 2d gauge theory as a 3d gauge theory on an interval with half-BPS boundary conditions {\it a la} \cite{Gadde:2013wq,Gadde:2013sca}.
For example, a 2d $(0,1)$ sigma-model on a group manifold $SU(2)$ can be realized in this way by imposing Dirichlet boundary conditions on both sides of the interval in 3d $\CN=1$ super-Chern-Simons theory \cite{Gaiotto:2019asa}.
If, instead, we impose Neumann boundary conditions on both sides, then the gauge symmetry is preserved and we obtain a 2d $\CN=(0,1)$ gauge theory with gauge group $G = SU(2)$.
In this model, the gauge anomaly needs to be canceled, and the anomaly inflow from 3d at the two boundaries is $+k+1$ and $-k+1$, where $k$ is the Chern-Simons level and $+1$ is the contribution of fermions (which has opposite sign compared to the case of Dirichlet boundary condition \cite{Gadde:2013sca}). Note, that the Chern-Simons term contributes with opposite signs since the relative orientation of the boundaries is reversed.

When $k=1$, several nice things happen. First, this is the smallest value of $k$ for which 3d $\CN=1$ super-Chern-Simons does not break supersymmetry; in fact, in the IR it is believed to have only one quantum ground state~\cite{Witten:1999ds}. Secondly, at $k=1$ the anomaly at one boundary vanishes and the anomaly at the second boundary is $+1+1=2$. It can be canceled by 4 fundamental $(0,1)$ scalars, {\it i.e.} $N_f = 2$ fundamental $(0,2)$ chirals. Not surprisingly, this is the same anomaly-free matter content of 2d $(0,1)$ SQCD with $N_c=2$ that we found earlier. And, since the physics on one of the boundaries is trivial and the bulk $SU(2)_1$ theory has a unique gapped SUSY vacuum, in the IR this sandwich of 3d-2d theories flows to a theory of gauge-invariant operators on the second boundary, {\it i.e.} NLSM with target $\C^4/ SU(2) \cong \R^5$ that we found earlier via a different but similar route.\footnote{A generalization to theories \eqref{NcNfquiver} with arbitrary values of $N_f = N_c$ is straightforward and involves 3d $\CN=1$ super-Chern-Simons $SU(N_c)_k$ at level $k = \frac{N_c}{2}$. At this special value of $k$, the anomalies at the two boundaries are $k + \frac{N_c}{2} = N_c$ and $-k + \frac{N_c}{2}=0$. In the IR, the bulk 3d theory has only one gapped SUSY vacuum and the effective 2d $(0,1)$ physics is described by mesons on a ``non-trivial boundary'' that parametrize~\eqref{noncomptarget}.}

In order to gain a new perspective on 2d $(0,1)$ SQCD \eqref{NcNfquiver}, we need a 3d theory sandwiched by two half-BPS boundary conditions such that $N_f$ fundamental matter fields also have a 3d origin. The most natural 3d theory that meets these criteria is a 3d $\CN=1$ vector multiplet coupled to $N_f = N_c$ copies of 3d $\CN=2$ chiral multiplets in the fundamental representation of $G=SU(N_c)$. Equivalently, this theory can be viewed as a 3d $\CN=2$ theory with $N_f = N_c$ fundamental chirals, from which half of the gauginos have been removed. However, the IR physics of such 3d $\CN=1$ theories does not appear to be understood at the time of this writing.

\section*{Acknowledgments}
We would like to thank Francesco Benini, Mykola Dedushenko, Shiraz Minwalla, Cumrun Vafa, Edward Witten for fruitful discussions.
The work of S.G.~is supported by the U.S. Department of Energy, Office of Science, Office of High Energy Physics, under Award No. DE-SC0011632, and by the National Science Foundation under Grant No. NSF DMS 1664240. The work of D.P.~is supported by NSF Grant DMS-1440140 while in residence at the Mathematical Sciences Research Institute in Berkeley, California during the Fall 2019 semester, and by the center of excellence grant ``Center for Quantum Geometry of Moduli Space" from the Danish National Research Foundation (DNRF95). 

\appendix

\section{Cohomology and characteristic classes of real Grassmannians}
\label{app:coh-grassmannian}

\subsection{Mod-2 cohomology and Stiefel--Whitney classes}

The cohomology of the Grassmannian can be determined from their coset descriptions. For the Grassmannian of unoriented $n$-planes in $\R^N$ we have
\begin{equation}
    Gr(n,N)=\cfrac{O(N)}{O(n)\times O(N-n)}.
\end{equation}
It follows that the Grassmannian is the fiber of the following fibation of the classifying spaces:
\begin{equation}
    Gr(n,N)\longrightarrow BO(n)\times BO(N-n) \longrightarrow BO(N).
    \label{grassmannian-fibration}
\end{equation}
In other words, $Gr(n,N)$ can be understood as the classifying space for pairs of rank-$n$ and rank-$(N-n)$ bundles whose direct sum is a trivial rank-$N$ bundle. To determine the cohomology of the Grassmannian, it is then enough to know the cohomology of $BO(n)$ and the action of the map $BO(n)\times BO(m)\rightarrow BO(n+m)$ corresponding to Whitney sum of real vector bundles. In the case of $\Z_2$ coefficients, the cohomology of $BO(n)$ is well known to be freely generated by Stiefel--Whitney classes,
\begin{equation}
    H^*(BO(n),\Z_2)\cong \Z_2[w_1,w_2,\ldots,w_n],\qquad \deg w_i=i.
\end{equation}
The total Stiefel--Whitney class $w:=1+\sum_{i=1}^n w_i$ of a Whitney sum of bundles is the product of their total Stiefel--Whitney classes. This gives the following simple explicit description of the mod-2 cohomology of a real Grassmannian:
\begin{equation}
    H^*(Gr(n,N),\Z_2)\cong \frac{\Z_2[w_1,w_2,\ldots,w_n,\bar{w}_1,\bar{w}_2,\ldots,\bar{w}_{N-n}]}{(1+w_1+w_2+\ldots+w_n)(1+\bar{w}_1+\bar{w}_2+\ldots+\bar{w}_{N-n})=1}.
\end{equation}
By construction, the classes $w_i$ and $\bar{w}_i$ can be identified with the Stiefel--Whitney classes of the tautological and orthogonal bundles respectively. That is
\begin{equation}
 w_i=w_i(S),\qquad \bar{w}_i=w_i(Q).
\end{equation}
The Stiefel--Whiteny classes of the tangent bundle can be determined from the relation $TX\cong S\otimes Q$. Although the Stiefel--Whitney classes of the tensor product can be determined in terms of Stiefel--Whitney classes of the factors using the splitting principle, the explicit expressions become quite complicated already in degrees 3 and 4. However they simplify significantly if the bundles are the same. Namely, for any rank $m$ real vector bundle $V$ we have
\begin{equation}
    \begin{array}{rl}
        w_1(V\otimes V) =& 0, \\
        w_2(V\otimes V) =& (m-1)\,w_1^2(V), \\
        w_3(V\otimes V) =& 0, \\
        w_4(V\otimes V)=& \frac{(m-1)(m-2)}{2}\,w_1^4(V)+m\,w_2^2(V), \\
        \vdots &
    \end{array}
\end{equation}
Then, using $S\otimes Q \oplus S\otimes S=S^{N_1}$ we have
\begin{multline}
    w(TGr(n,N))=w(S\otimes Q)=\frac{w(S)^{N}}{w(S\otimes S)}=\\
    \left(1+N(w_1+w_2+w_3+w_4+\ldots)+\frac{N(N-1)}{2}(w_1^2+w_2^2+\ldots)+\right. \\
    \left.\frac{N(N-1)(N-2)}{3!}(w_1^3+\ldots)+\frac{N(N-1)(N-2)(N-3)}{4!}(w_1^4+\ldots)+\ldots\right)/\\
    \left(1+(n-1)w_1^2+nw_2^2+\frac{(n-1)(n-2)}{2}w_1^4+\ldots\right)
\end{multline}
where we have kept only elements up to degree 4.

The case of the oriented Grassmannian
\begin{equation}
    \tilde{Gr}(n,N)\cong \frac{SO(N)}{SO(n)\times SO(N-n)}
\end{equation}
can be deduced simply by setting $w_1=\bar{w}_1=0$ in the above formulas, due to the fact that 
\begin{equation}
        H^*(BSO(n),\Z_2)\cong \Z_2[w_2,\ldots,w_n],\qquad \deg w_i=i.
\end{equation}

\subsection{Rational cohomology and rational Pontryagin classes}

The cohomology with rational coefficients can be obtained by a similar approach, starting from the fact that
\begin{equation}
    H^*(BO(n),\Q)\cong \Q[p_1,p_2,\ldots,p_n],\qquad \deg p_i=4i.
\end{equation}
This leads to the following description of the cohomology of the Grassmannian,
\begin{equation}
    H^*(Gr(n,N),\Q)\cong \frac{\Q[p_1,p_2,\ldots,p_n,\bar{p}_1,\bar{p}_2,\ldots,\bar{p}_{N-n}]}{(1+p_1+p_2+\ldots+p_n)(1+\bar{p}_1+\bar{p}_2+\ldots+\bar{p}_{N-n})=1},
\end{equation}
where the generators are Pontryagin classes of the tautological and orthogonal bundles, i.e.~$p_i=p_i(S)$ and $\bar{p}_i=p_i(Q)$. The total rational Pontryagin class of the tangent bundle can be obtained again as
\begin{equation}
     p\left(TGr(n,N)\right)=p(S\otimes Q)=\frac{p(S)^{N}}{p(S\otimes S)}=
     \frac{1+Np_1+\ldots}{1+2np_1+\ldots}=1+(N-2n)p_1+\ldots
\end{equation}
and using the explicit formulas for the Pontryagin classes of the tensor product. The rational cohomology of the oriented Grassmannian $\tilde{Gr}(n,N)$ is the same.


\subsection{Integral cohomology}

The integral cohomology of a real Grassmannian is quite involved (unlike the case of the complex Grassmannian), but in principle can be still obtained from the fibration (\ref{grassmannian-fibration}) and the knowledge of the integral cohomology of $BO(n)$ (or $BSO(n)$ in the oriented case) as well as the pullback of the Whitney sum map $BO(n)\times BO(N-n)\rightarrow BO(N)$ (or $BSO(n)\times BSO(N-n)\rightarrow BSO(N)$ in the oriented case) which were described in \cite{brown1982cohomology}. When $n$ is sufficiently large, the cohomology of $BO(n)$ in low degrees is generated by Pontryagin classes as well as the images of Stiefel--Whitney classes under the Bockstein homomorphism 
\begin{equation}
    \delta:\;H^*(BO(n),\Z_2)\longrightarrow H^{*+1}(BO(n),\Z)
\end{equation}
associated to the short exact sequence of coefficients
\begin{equation}
    \Z\stackrel{2\cdot}{\longrightarrow} \Z \stackrel{\bmod 2}{\longrightarrow} \Z_2.
\end{equation}
The generators are subject to the certain relations. In low degrees, the cohomology groups and the independent generators read
\begin{equation}
\begin{tabular}{c|c|c}
$i$ & $H^i(BO(n),\Z)$ & generators \\
\hline
     0 & $\Z$ & 1  \\
     1 & 0 &  \\
     2  & $\Z_2$ & $\delta(w_1)$ \\
     3 & $\Z_2$ & $\delta(w_2)$ \\
     4 & $\Z\times \Z_2^2$ & $p_1,\,\delta(w_1w_2),\;(\delta(w_1))^2$ \\
     \ldots & \ldots & \ldots
\end{tabular}
\end{equation}
The cohomology of $Gr(n,N)$ will be the same in low degrees when $n$ and $N$ are sufficiently low. Similarly, the cohomology of the oriented Grassmannian $\tilde{Gr}(n,N)$ in low degrees coincides with the cohomology of $BSO(n)$:
\begin{equation}
\begin{tabular}{c|c|c}
$i$ & $H^i(BSO(n),\Z)$ & generators \\
\hline
     0 & $\Z$ & 1  \\
     1 & 0 &  \\
     2  & 0 &  \\
     3 & $\Z_2$ & $\delta(w_2)$ \\
     4 & $\Z$ & $p_1$ \\
     \ldots & \ldots & \ldots
\end{tabular}
\end{equation}
The details of the cohomology and the explicit expression for the integral Pontryagin classes of the tangent bundle will not be actually important for us in this paper. This is because the torsion part contains only order-2 elements and, therefore, it will be sufficient to work with mod-2 and rational cohomologies.

\bibliographystyle{JHEP}
\bibliography{tri-bib}

\end{document}